%% file: main.tex
\documentclass[a4paper,fleqn,usenatbib]{mnras}
 
\usepackage[T1]{fontenc}
\usepackage{ae,aecompl}
\usepackage[symbol]{footmisc}

\usepackage{graphicx}	
\usepackage{amsmath}	
\usepackage{amssymb}	
\usepackage{hyperref}
\usepackage{xcolor}
\usepackage{siunitx}
\usepackage{subfig}
\usepackage{soul}

\newcommand{\sub}[2]{#1_{\mathrm{#2}}}

\DeclareSIUnit{\nucleon}{nucleon}
\DeclareSIUnit{\erg}{erg}
\DeclareSIUnit{\mevnuc}{\MeV\per\nucleon}
\DeclareSIUnit{\Funits}{\erg\per\second\per\square\cm}
\DeclareSIUnit{\flunits}{\erg\per\square\cm}
\DeclareSIUnit{\yunits}{\gram\per\square\cm}
\DeclareSIUnit{\msun}{M_\odot}

\newcommand{\saxj}{SAX J1808.4$-$3658}
\newcommand{\gs}{GS 1826$-$238}
\newcommand{\fouru}{4U 1820$-$30}

\newcommand{\kepler}{\textsc{Kepler}}
\newcommand{\mesa}{\textsc{MESA}}
\newcommand{\shiva}{\textsc{SHIVA}}
\newcommand{\settle}{\textsc{Settle}}
\newcommand{\emcee}{\textsc{emcee}}

\newcommand{\python}{\textsc{python}}

\newcommand{\nw}[1]{\sub{#1}{k}}
\newcommand{\gr}[1]{\sub{#1}{g}}

\newcommand{\dt}{\Delta t}
\newcommand{\brate}{\nu}
\newcommand{\Eb}{\sub{E}{b}}
\newcommand{\fluence}{\sub{f}{b}}
\newcommand{\Fper}{\sub{F}{p}}
\newcommand{\Fpeak}{\sub{F}{peak}}
\newcommand{\Fedd}{\sub{F}{Edd}}
\newcommand{\Lper}{\sub{L}{p}}
\newcommand{\Lpeak}{\sub{L}{peak}}
\newcommand{\Ledd}{\sub{L}{Edd}}

\newcommand{\obslhood}{0}
\newcommand{\obssymb}{\infty}
\newcommand{\obs}[1]{\sub{#1}{\obssymb}}
\newcommand{\obssub}[2]{\sub{#1}{#2, \obssymb}}

\newcommand{\qb}{\sub{Q}{b}}
\newcommand{\mdot}{\dot{m}}
\newcommand{\Mdot}{\dot{M}}
\newcommand{\mdotedd}{\sub{\mdot}{Edd}}
\newcommand{\Mdotedd}{\sub{\Mdot}{Edd}}
\newcommand{\qnuc}{\sub{Q}{nuc}}
\newcommand{\cno}{Z_\mathrm{CNO}}
\newcommand{\hyd}{X_0}
\newcommand{\xib}{\sub{\xi}{b}}
\newcommand{\xip}{\sub{\xi}{p}}
\newcommand{\xiratio}{\xip / \xib}
\newcommand{\db}{d \sqrt{\xib}}

\newcommand{\qbn}[1]{Q_{\mathrm{b}, #1}}
\newcommand{\mdotn}[1]{\mdot_{#1}}
\newcommand{\Mdotn}[1]{\Mdot_{#1}}

\newcommand{\posterior}{p(\theta | D)}
\newcommand{\likelihood}{p(D | \theta)}
\newcommand{\prior}{p(\theta)}

\defcitealias{galloway_periodic_2004}{G04}
\defcitealias{heger_models_2007}{H07}
\defcitealias{meisel_consistent_2018}{M18}
\defcitealias{galloway_thermonuclear_2017}{G17}

\title[Multi-epoch X-ray burst modelling]{Multi-epoch X-ray burst modelling: MCMC with large grids of 1D simulations}

\author[Z. Johnston et al.]{
Zac Johnston,$^{1,2,3}$\thanks{E-mail: zacjohn@msu.edu}
Alexander Heger,$^{1,2,4,5}$
Duncan K. Galloway$^{1,2}$
\\
$^{1}$School of Physics and Astronomy, Monash University, Victoria 3800, Australia\\
$^{2}$Monash Centre for Astrophysics, Monash University, Victoria 3800, Australia\\
$^{3}$Department of Physics and Astronomy, Michigan State University, East Lansing, Michigan 48824, USA\\
$^{4}$Department of Astronomy,
Shanghai Jiao Tong University, Shanghai 200240, China\\
$^{5}$School of Physics and Astronomy, University of Minnesota, Minneapolis,
Minnesota 55455, USA\\
}
\date{Accepted XXX. Received YYY; in original form ZZZ}
\pubyear{2020}

\begin{document}
\label{firstpage}
\pagerange{\pageref{firstpage}--\pageref{lastpage}}
\maketitle

\input{misc/abstract}


\section{Introduction}
\label{sec:intro}
\input{sections/introduction}

\section{Methods}
\label{sec:method}
\input{sections/method}

\section{Results}
\label{sec:results}
\input{sections/results}

\section{Discussion}
\label{sec:discussion}
\input{sections/discussion}

\section{Conclusion}
\label{sec:conclusion}
\input{sections/conclusion}

\section*{Acknowledgements}
\input{misc/acknowledgements}

\section*{ORCID iDs}
Zac Johnston \href{https://orcid.org/0000-0003-4023-4488}{https://orcid.org/0000-0003-4023-4488}\\
Alexander Heger \href{https://orcid.org/0000-0002-3684-1325}{https://orcid.org/0000-0002-3684-1325}\\
Duncan K. Galloway \href{https://orcid.org/0000-0002-6558-5121}{https://orcid.org/0000-0002-6558-5121}

\bibliographystyle{mnras}
\bibliography{main} 

\appendix

\section{Model preheating}
\label{sec:appendix_preheating}
\input{appendix/preheating}

\section{GR corrections}
\label{sec:appendix_gr}
\input{appendix/gr}

\section{Model data}
\label{sec:appendix_data}
\input{appendix/data}

\bsp	
\label{lastpage}
\end{document}

%% file: misc/abstract.tex
\begin{abstract}
Type-I X-ray bursts are recurring thermonuclear explosions on the surface of accreting neutron stars.
Matching observed bursts to computational models can help to constrain system properties, such as the neutron star mass and radius, crustal heating rates, and the accreted fuel composition, but
systematic parameter studies to date have been limited.
We apply Markov chain Monte Carlo methods to 1D burst models for the first time, and obtain system parameter estimations for the `Clocked Burster', GS 1826$-$238, by fitting multiple observed epochs simultaneously.
We explore multiple parameters which are often held constant, including the neutron star mass, crustal heating rate, and hydrogen composition.
To improve the computational efficiency, we precompute a grid of 3840 \textsc{Kepler} models -- the largest set of 1D burst simulations to date -- and by interpolating over the model grid, we can rapidly sample burst predictions.
We obtain estimates for a CNO metallicity of $Z_\mathrm{CNO} = 0.010^{+0.005}_{-0.004}$, a hydrogen fraction of $X_0 = 0.74^{+0.02}_{-0.03}$, a distance of $d \sqrt{\xi_\mathrm{b}} = 6.5^{+0.4}_{-0.6}\, \mathrm{kpc}$ , and a system inclination of $i = {69^{+2}_{-3}}^{\circ}$.
\end{abstract}

%% file: sections/introduction.tex
Type-I thermonuclear X-ray bursts are recurring flashes observed from accreting neutron stars \citep[for reviews, see][]{lewin_x-ray_1993, strohmayer_new_2006, galloway_thermonuclear_2017-1}.
In the host low-mass X-ray binary (LMXB) systems, a neutron star accretes material from a companion star with a mass of $M \lesssim \SI{1}{\msun}$, which accumulates as a $\sim \SI{10}{m}$ envelope on the neutron star surface.
Under the weight of accreting material, the base of the envelope is compressed by the extreme surface gravity of $g \sim \SI{e14}{cm.s^{-2}}$ to the point of thermonuclear runaway.
Within seconds, the layer is heated to $\sim \SI{e9}{K}$, generating a burst of X-rays before cooling to background levels over the following seconds to minutes.
New fuel is accreted on top of the `ashes', and the cycle repeats.

X-ray bursts have been the target of numerical calculations since the 1970s \citep[e.g.,][]{joss_helium-burning_1978, taam_thermonuclear_1979}, and their diverse behaviour has been studied with a variety of computational models \citep[e.g.,][]{fujimoto_theory_1987, woosley_models_2004, keek_superburst_2012}.
By exploring model parameters and comparing the predictions with observations, the neutron star system properties can be inferred \citep[e.g.,][]{cumming_models_2003, galloway_periodic_2004, keek_thermonuclear_2017, johnston_simulating_2018}.

One-dimensional (1D) burst codes are the best tools currently available for this purpose.
With adaptive nuclear reaction networks and treatments for convective transport \citep[e.g.,][]{woosley_models_2004}, they can produce detailed simulations of burst energetics not possible in semi-analytic or one-zone models.
Multi-dimensional burst simulations are also under active development, but computational costs limit the calculations to $\lesssim \SI{1}{s}$ of simulation time \citep[e.g.,][]{zingale_comparisons_2015, cavecchi_fast_2016}.

Nevertheless, targeted parameter explorations using 1D models have been relatively limited.
Small sets of models are typically used, and many parameters, such as the gravity, fuel composition, and crustal heating, are often held constant.
Due to the relatively unexplored parameter space, obtaining robust constraints on system properties is difficult.

To encourage modelling efforts, \citet[][hereafter, G17]{galloway_thermonuclear_2017} presented a set of standardised burst observations.
Their reference data set included three epochs of bursts from \gs{}, famously dubbed the `clocked burster' \citep[e.g.,][]{ubertini_bursts_1999}.
The system's reliability has made it a popular target for modelling  \citep[e.g.,][]{galloway_periodic_2004, heger_models_2007}, and particularly for the study of the nuclear \textit{rp}-process \citep[e.g.,][]{schatz_rp-process_1998, fisker_explosive_2008}.
The first study to make use of the \citetalias{galloway_thermonuclear_2017} data set was \citet[][hereafter, M18]{meisel_consistent_2018}, who performed the first extended comparison of \mesa{} burst models to \gs.
\citetalias{meisel_consistent_2018} demonstrated the benefit of fitting multiple epochs by ruling out parameter combinations which otherwise agreed with individual epochs.

\gs{} was discovered as a transient source with the \textit{Ginga} X-ray telescope in 1988 \citep{makino_gs_1988}, and X-ray bursts were later discovered in 1997 \citep{ubertini_gs_1997, ubertini_bursts_1999}.
The system orbital period is not precisely known, but is thought to be roughly \SI{2}{h} \citep{homer_evidence_1998}, implying a hydrogen-rich mass donor, consistent with the long-tailed bursts observed \citep{int_zand_long_2009}.
Despite the popularity of \gs{} for modelling, ambiguity persists regarding the system properties.
For example, \citet[][hereafter, G04]{galloway_periodic_2004} modelled bursts observed between 1997 and 2002 using a semi-analytic ignition model \citep[\settle, first used in][]{cumming_rotational_2000}.
They reported that an accreted CNO mass fraction of $\cno = 0.001$ best reproduced the trend of recurrence time, $\dt$, versus accretion rate, $\Mdot$, but that the observed $\alpha$ values (the ratio of integrated persistent flux to integrated burst flux) were only consistent with a higher metallicity of $\cno = 0.02$.
Using \kepler{}, \citet[][hereafter, H07]{heger_models_2007} found good lightcurve agreement\footnote{we note that these models were discovered to use inadvertently large opacities;
see \S~\ref{subsec:method_kepler}} for $\cno = 0.02$, and
\citetalias{meisel_consistent_2018} found agreement for both $\cno = 0.01$ and $0.02$ using \mesa{}.
The accretion rates are typically inferred to be in the range $\Mdot =$ 0.05--0.08 $\Mdotedd$ \citep{heger_models_2007, galloway_thermonuclear_2008, galloway_thermonuclear_2017}, where $\Mdotedd = \SI{1.75e-8}{\msun.yr^{-1}}$ is the fiducial rate for a $\SI{1.4}{\msun}$ neutron star under Newtonian gravity.
On the other hand, \citetalias{meisel_consistent_2018} reported improved model fits using twice as large accretion rates of $\Mdot =$ 0.1--0.17 $\Mdotedd$.

The inconclusive estimates are, we suggest, partly due to the limited parameter explorations to date, in addition to degeneracies between the model predictions.
For example, the metallicity is often fixed at $\cno = 0.02$, with an accreted hydrogen fraction of $\hyd = 0.7$.
\citetalias{galloway_periodic_2004} and \citetalias{heger_models_2007} used fixed values for base heating (i.e., the flux emerging from the crust) of $\qb = 0.1$ and $\SI{0.15}{\mevnuc}$, respectively, whereas \citetalias{meisel_consistent_2018} considered $\qb = 0.1$, 0.5, and $\SI{1.0}{\mevnuc}$.
Both \citetalias{heger_models_2007} and \citetalias{meisel_consistent_2018} assumed a fixed neutron star mass of $M = \SI{1.4}{\msun}$ and a radius of $R = \SI{11.2}{km}$, whereas \citetalias{galloway_periodic_2004} assumed $M = \SI{1.4}{\msun}$ and $R = \SI{10}{km}$.
The earlier estimates for $\Mdot$ did not account for the possible effect of anisotropic emission (\S~\ref{subsec:method_free_params}), which is dependent on the system inclination and disc morphology \citep{fujimoto_angular_1988}.
Using the disc models of \citet{he_anisotropy_2016}, \citetalias{meisel_consistent_2018} inferred an approximate inclination of 65--80$^\circ$, suggesting that the X-ray emission is preferentially beamed \textit{away} from the line of sight, allowing for larger $\Mdot$.
To fully account for the complex dependencies between these model parameters and predictions, a more comprehensive analysis is required.

Markov chain Monte Carlo (MCMC) methods are algorithms capable of sampling complex probability distributions \citep[for a comprehensive introduction, see][]{mackay_information_2003}.
The use of Bayesian statistics in astrophysics has seen a rapid expansion in recent years, but its application to X-ray burst modelling has been minimal.
Most recently, \citet{goodwin_bayesian_2019} applied MCMC methods to the semi-analytic burst code, \settle{}, to model bursts from the transient accretor, \saxj{}.
This system is also included in the \citetalias{galloway_thermonuclear_2017} data set, as an example of helium bursts triggered during an accretion event.
Pairing a semi-analytic model with MCMC is beneficial due to the computational speed required for drawing thousands of sequential samples.
By contrast, 1D burst models can take several days to compute, and are, on their own, unsuitable for MCMC methods.

As we show here for the first time, this computational barrier can be overcome with the use of pre-compiled model grids.
For burst properties that vary smoothly over the model parameters, interpolation can be used to sample bursts between existing models with little computational cost.
We present the first application of MCMC methods to large grids of 1D burst models.
By constructing a grid of \num{3840} \kepler{} simulations, we are able to rapidly sample burst properties across twelve parameters.
Using the data set from \citetalias{galloway_thermonuclear_2017}, we fit three epochs of burst data simultaneously, and obtain probability distributions for the system parameters of \gs{}.

In Section~\ref{sec:method}, we describe the \kepler{} code and its recent updates, the epoch data used, the construction and interpolation of the model grid, and the setup of the MCMC model.
In Section~\ref{sec:results}, we describe the model results, the posterior distributions, the predicted burst properties, and lightcurve comparisons.
In Section~\ref{sec:discussion}, we discuss and compare the parameter estimates to previous works, discuss the limitations of the model, and describe potential improvements to the model.
In Section~\ref{sec:conclusion}, we provide concluding remarks and the future outlook.

%% file: sections/method.tex

\begin{table*}
    {\centering
    \caption{The observed burst data from three epochs of \gs, adapted from the reference set of \citetalias{galloway_thermonuclear_2017}.
    As an additional system constraint, we have included the Eddington flux, $\Fedd$, taken from the peak of a PRE burst observed in 2014 June \citep{chenevez_soft_2016}.
    To estimate the average recurrence time $\dt$, \citetalias{galloway_thermonuclear_2017} collected multiple bursts from each epoch, from which we have obtained the burst rate $\brate$.
    They then extracted average lightcurves, from which the peak flux, $\Fpeak$, and fluence, $\fluence$, could be determined.
    The persistent flux $\Fper$ was averaged for each epoch, and the values listed here have incorporated the bolometric corrections estimated by \citetalias{galloway_thermonuclear_2017}.}
    \label{tab:obsdata}
    \begin{tabular}{cccccc}
        \hline
        \hline
        Epoch &  $\brate$         & $\Fpeak$              & $\fluence$             & $\Fper$               & $\Fedd$    \\
              & ($\si{day^{-1}}$) & $(\SI{e-9}{\Funits})$ & $(\SI{e-6}{\flunits})$ & $(\SI{e-9}{\Funits})$ & $(\SI{e-9}{\Funits})$ \\
        \hline
        1998 Jun & $4.67  \pm 0.06$  & $30.9 \pm 1.0$ &  $1.102 \pm 0.011$ & $2.108 \pm 0.015$ & -- \\
        2000 Sep & $5.746 \pm 0.014$ & $29.1 \pm 0.5$ & $1.126  \pm 0.016$ & $2.85  \pm 0.03$  & -- \\
        2007 Mar & $6.799 \pm 0.008$ & $28.4 \pm 0.4$ & $1.18   \pm 0.04$  & $3.27  \pm 0.04$  & -- \\
        2014 Jun & --                & --             & --                 & --                & $40 \pm 3$ \\
        \hline
    \end{tabular}\\
    }
\end{table*}

\begin{figure}
    \centering
     \subfloat{\includegraphics[width=\linewidth]{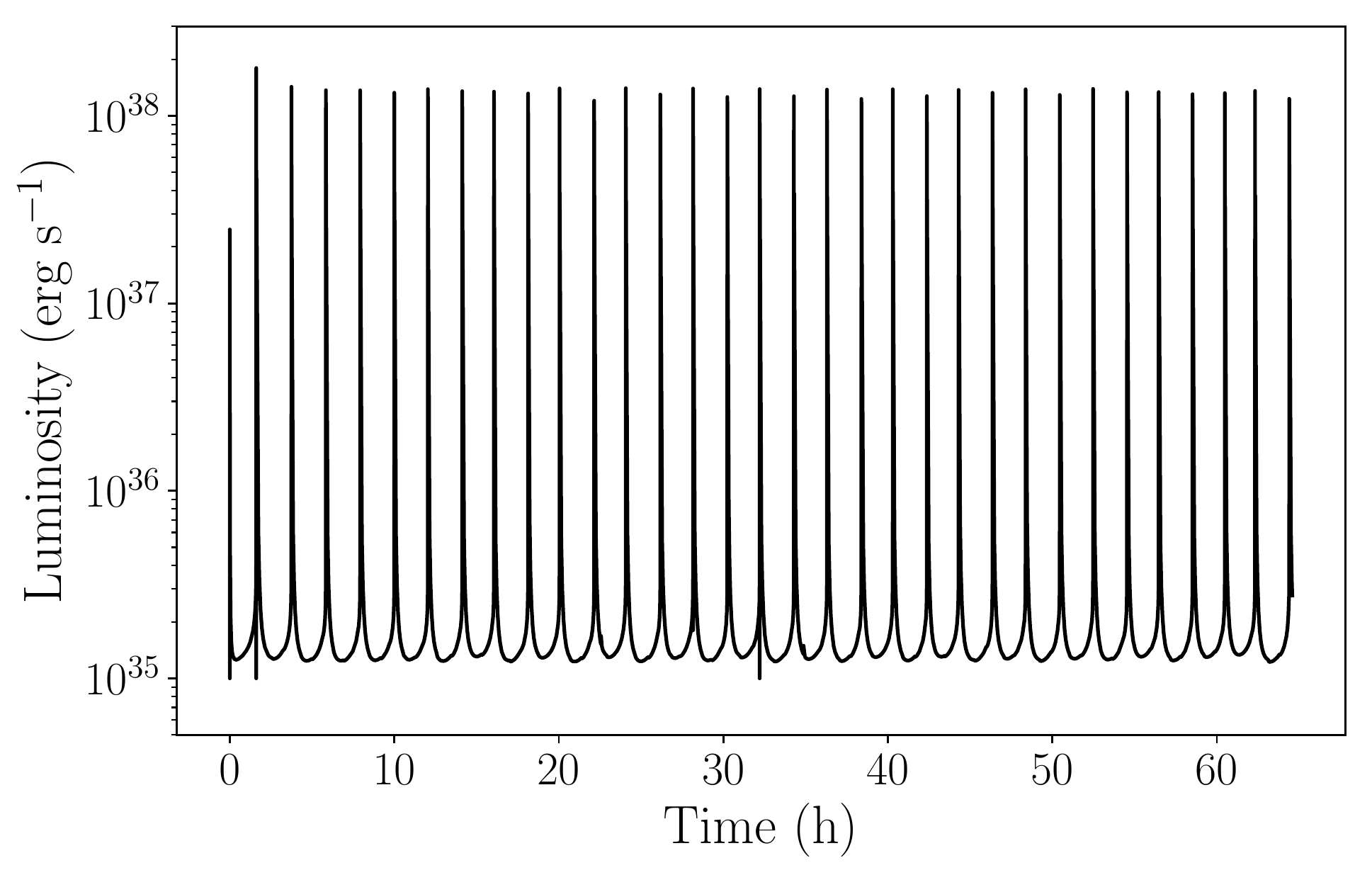}}

     \subfloat{\includegraphics[width=\linewidth]{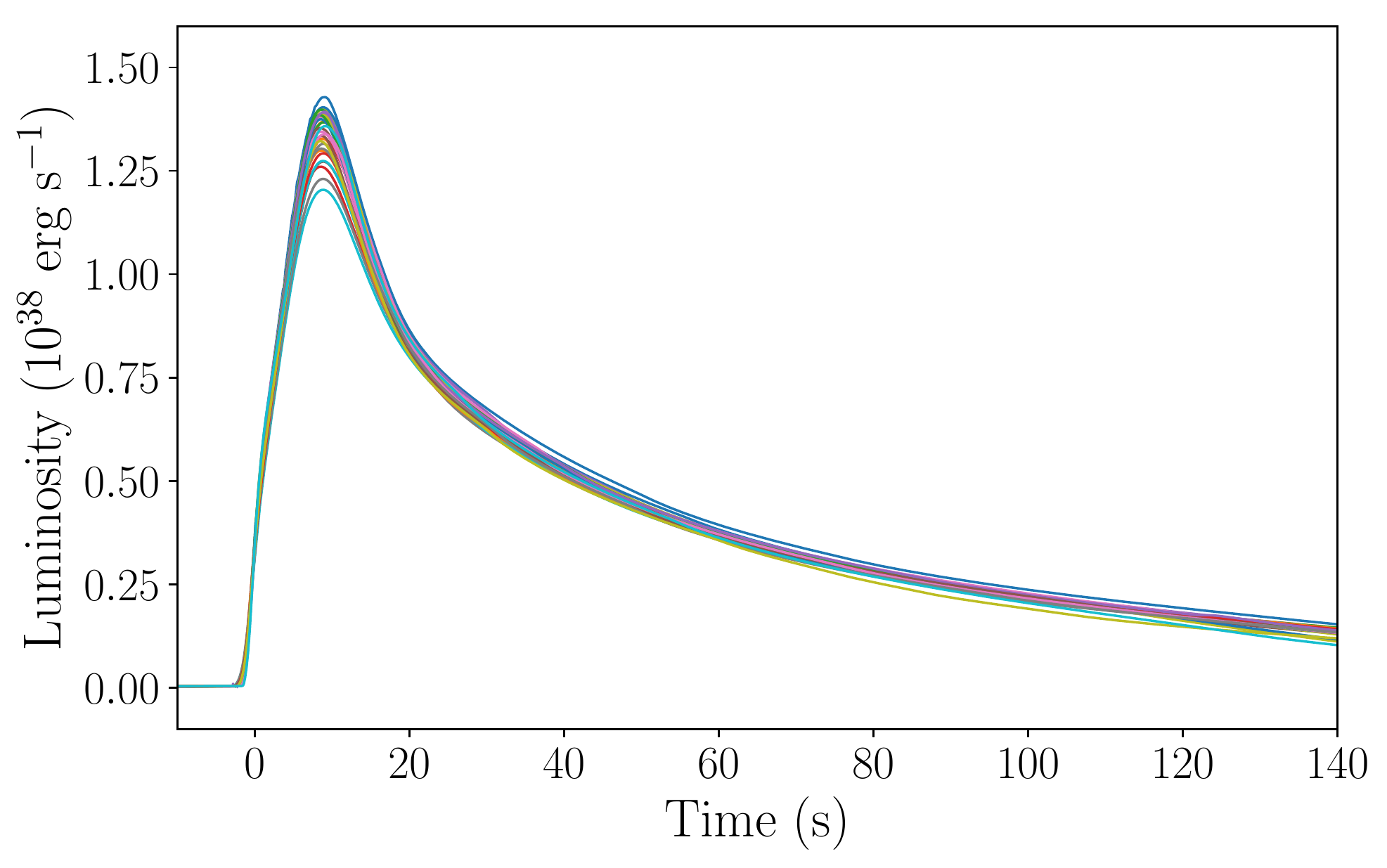}}
    \caption{An example burst train simulated using \kepler{} (upper panel), and the extracted and stacked burst lightcurves (lower panel) from which average properties can be calculated (\S~\ref{subsec:method_extracting}).
    }
    \label{fig:sequence}
\end{figure}

\subsection{An update on \kepler}
\label{subsec:method_kepler}
\kepler{} \citep{weaver_presupernova_1978} is a one-dimensional (1D) hydrodynamics code capable of simulating a variety of regimes in stellar evolution and explosive nucleosynthesis \citep[e.g.,][]{woosley_evolution_2002, menon_quest_2017}.
It has prominently been used for modelling X-ray bursts, reproducing observed behaviour, including burst energetics, recurrence times, and lightcurves \citep{woosley_models_2004, heger_models_2007, keek_superburst_2012, lampe_influence_2016}.
Because \kepler{} has steadily been modified and improved over time, some descriptions in earlier works are now out of date.
We here briefly summarise notable changes to the code and model setup.

To aid reproducibility and comparisons to other burst models, we used V2.2 of JINA REACLIB, the public database of nuclear reaction rates\footnote{\url{https://jinaweb.org/reaclib/db/}} \citep{cyburt_jina_2010}.

\kepler{} burst models published prior to \citet{johnston_simulating_2018} used a setup file which erroneously multiplied the opacities by a factor of ${\approx} 1.5$.
The original intent was to approximate the time-dilation effects of general relativity (GR; Appendix~\ref{sec:appendix_gr}) by artificially slowing down thermal transport.
While the idea was soon abandoned and removal of the factor was intended, it mistakenly remained in the setup file which was adapted for subsequent studies.
This error was discovered\footnote{by Adam Jacobs, Michigan State University, pers.\ comm.\ } and amended for the models presented in \citet{johnston_simulating_2018}.
The boosted opacities reduced the thermal conductivity, resulting in an artificially hotter envelope and shorter burst recurrence times.
This discrepancy likely explains why comparisons from other burst codes, for example \mesa{} \citepalias{meisel_consistent_2018} and \shiva{} \citep{jose_hydrodynamic_2010}, reported longer recurrence times than the equivalent \kepler{} models.
This issue should be kept in mind when making comparisons to previous \kepler{} models \citep[e.g.,][]{heger_models_2007,lampe_influence_2016}.

During the setup phase of the model envelope, before accretion and nuclear reactions are switched on, the thermal profile is initialised near equilibrium, in order to minimise simulation `burn-in'.
In previous \kepler{} studies (and to our knowledge, all other burst studies in the literature), the only included heat source was the base heating parameter, $\qb \approx \SI{0.15}{MeV.nucleon^{-1}}$, as a boundary condition at the base of the model domain, at a column depth of $y \approx \SI{1e12}{\yunits}$.
This parameter represents the flux emerging from heating processes in the crust, which lies below the model domain at $y \gtrsim \SI{1e13}{\yunits}$.
The heat generated in the envelope by nuclear reactions, $\qnuc \approx \SI{5}{\mevnuc}$, around $y \sim \SI{e8}{\yunits}$, was assumed to largely escape the surface, and was neglected from the setup calculations.
Minor heating of the deeper ocean, once the full nuclear simulation began, was expected to stabilise after the initial few bursts.
For example, \citet{woosley_models_2004} discarded the first three bursts from analysis to address this `thermal inertia', in addition to the related effect of `chemical inertia' (\S~\ref{subsec:method_extracting}).

During testing, however, we discovered that nuclear heating can indeed alter the thermal profile enough to influence burst ignition.
Models which do not account for $\qnuc$ during setup begin comparatively colder, producing a steadily-changing burst sequence as the envelope is heated by nuclear reactions towards a steady thermal state.
This burn-in period can persist for dozens of bursts -- much longer than previously assumed.
To address this issue, we added during setup a heat source of $\qnuc = \SI{5}{\mevnuc}$ at a depth of $y \approx \SI{8e7}{\yunits}$ with a Gaussian spread of $y \approx \SI{8e6}{\yunits}$.
This heat source is switched off once the full nuclear calculations begin (further detail is provided in Appendix~\ref{sec:appendix_preheating}).
The model burn-in was effectively eliminated, and the burst sequence was stabilised within the first few bursts, as was originally assumed.
Further study is still required to explore the optimal configuration of this `preheating', which is likely to depend on other model parameters, such as the composition, accretion rate, and crustal heating.

\subsection{Observed data}
\label{subsec:method_observations}
We used observations of bursts from \gs{} for three epochs: 1998 June, 2000 September, and 2007 March (Table~\ref{tab:obsdata}).
These observations were provided as part of a reference set for burst modelling by \citetalias{galloway_thermonuclear_2017}, using data from the MINBAR catalogue \citep{galloway_multi-instrument_2020}.
The epochs were selected by \citetalias{galloway_thermonuclear_2017} for their burst consistency and the availability of high-precision X-ray lightcurves from the \textit{RXTE} satellite.

We also included the peak bolometric flux of $\Fpeak = \SI{40 \pm 3 e-9}{\Funits}$, observed from a photospheric radius expansion (PRE) burst in 2014 \citep{chenevez_soft_2016}.
We have assumed that $\Fpeak$ corresponds to the local Eddington luminosity, $\Fedd$, for a mixed hydrogen/helium envelope (\S~\ref{subsec:method_observables}), and that $\Fedd$ is common to the 1998--2007 epochs.

\subsection{Model grid}
\label{subsec:method_grid}
To model the bursts of \gs{}, we computed a regular grid of 3840 \kepler{} simulations over five model parameters: the accretion rate, $\mdot$, the accreted hydrogen mass fraction $\hyd$, the accreted CNO-metallicity mass fraction, $\cno$, the base heating, $\qb$, and the surface gravitational acceleration, $g$.
Note that for the \kepler{} model parameters, we give the local accretion rate per unit area, $\mdot$, because the global accretion rate, $\Mdot = 4 \pi R^2 \mdot$, depends on the choice of $R$ (\S~\ref{subsec:method_free_params}).
The numerical parameters controlling zone resolution were held constant, following convergence tests to ensure consistent burst sequences.
The grid points for each parameter are listed in Table~\ref{tab:params}.

Following a trial parameter exploration, we chose a grid that approximately covered the observed recurrence times of $3 \lesssim \dt \lesssim \SI{6}{\hour}$.
The model grid represents over $\num{100000}$ CPU hours, and is the largest collection of 1D burst models to date, with the previous largest containing 464 \kepler{} models \citep{lampe_influence_2016}.

Each model generated a sequence of 30--40 bursts (Fig.~\ref{fig:sequence}).
Simulating a long sequence ensures a consistent burst train, and reduces the effect of model burn-in (\S~\ref{subsec:method_kepler}).
The entire grid contained a total of approximately $\num{138000}$ bursts.
Modelling such large collections of bursts has been made possible by improved CPU speeds, which have reduced the computational cost from $\approx \SI{24}{\hour}$ per burst in 2003 to $\approx \SI{1}{\hour}$ per burst.

\begin{table*}
    {\centering
    \caption{The parameters of the model grid.
    Every combination was simulated, forming a regular grid in 5 dimensions.
    All models used a preheating value of $\qnuc = \SI{5}{\mevnuc}$ (\S~\ref{subsec:method_kepler}).
    The accretion rate, $\mdot$, is the local rate per unit area, because the global accretion rate, $\Mdot$, is dependent on the choice of $R$ (\S~\ref{subsec:method_free_params}).
    The Eddington-limited accretion rate, $\mdotedd = \SI{8.775e4}{g.cm^{-2}.s^{-1}}$ (assuming $M = \SI{1.4}{\msun}$, $R = \SI{10}{km}$, and $X = 0.7$), is simply used as a common reference point between models, and was not corrected for the $g$ or $\hyd$ of each model.
    The values of $g$ correspond to Newtonian surface gravities for masses of $M = 1.4,\, 1.7,\, 2.0,$ and $\SI{2.6}{\msun}$, for a reference radius of $R = \SI{10}{km}$, but the actual mass and radius are free parameters (\S~\ref{subsec:method_free_params}).
    }
    \label{tab:params}
    \begin{tabular}{ccccc}
        \hline
        \hline
        Parameter & Name & Units & Grid Points  &  N  \\
        \hline
        $\mdot$ & Local accretion rate          & ($\mdotedd$)            & 0.06, 0.07, 0.08, 0.10, 0.12, 0.14, 0.16, 0.18  & 8  \\
        $\hyd$  & H mass fraction               & --                      & 0.64, 0.67, 0.70, 0.73, 0.76                    & 5  \\
        $\cno$  & CNO mass fraction             & --                      & 0.0025, 0.005, 0.0075, 0.0125, 0.02, 0.03       & 6  \\
        $\qb$   & Base heating                  & ($\si{\mevnuc}$)        & 0.0, 0.2, 0.4, 0.6                              & 4  \\
        $g$     & Surface gravity               & ($\SI{e14}{cm.s^{-2}}$) & 1.86, 2.26, 2.65, 3.45                          & 4  \\
        \hline
        \multicolumn{4}{r}{Total}  & 3840 \\
        \hline
    \end{tabular}\\
    }
\end{table*}

\begin{figure}
  \centering
 	\includegraphics[width=\linewidth]{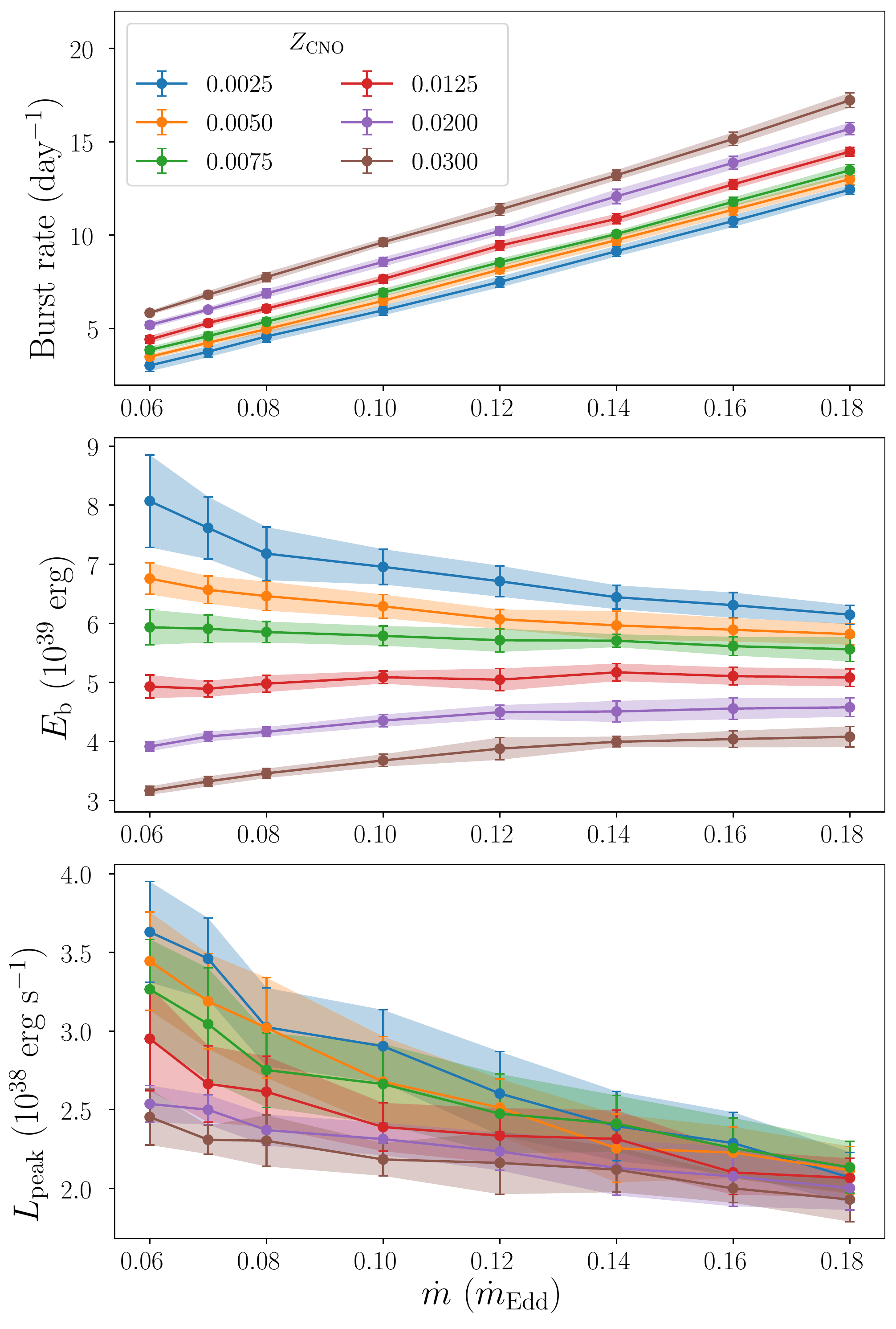}
	\caption{A subset of 48 \kepler{} models, from a total grid of 3840, illustrating the dependence of the burst properties on accretion rate, $\mdot$, and metallicity, $\sub{Z}{CNO}$.
    The subset is a grid slice through $\hyd = 0.64$, $\qb = \SI{0.2}{\mevnuc}$, and $g = \SI{2.26e14}{cm.s^{-2}}$.
    Each point was averaged from a sequence of 20--30 bursts (\S~\ref{subsec:method_extracting}), and the error bars are $1 \sigma$ standard deviations.
    The overall smooth and monotonic behaviour allows for the use of interpolation.
    The solid lines and shaded uncertainty regions were linearly interpolated between the models.
    Multivariate linear interpolation was used across all grid parameters for MCMC sampling.
    }
    \label{fig:grid}
\end{figure}

\subsection{Extracting model bursts}
\label{subsec:method_extracting}
Analysing bursts from \kepler{} models has been detailed previously \citep{woosley_models_2004, heger_models_2007, keek_multi-zone_2011, lampe_influence_2016}.
We briefly summarise our analysis routines, which are available as an open source \python{} package\footnote{\url{https://github.com/zacjohnston/pyburst/}}.

The procedure identified bursts from local maxima in the model lightcurve, from which the the individual lightcurves were extracted and analysed (Fig.~\ref{fig:sequence}).
The peak luminosity, $\Lpeak$, was taken from the lightcurve maximum.
The burst energy, $\Eb$, was the time-integrated luminosity over the lightcurve.
The recurrence time, $\dt$, was the time since the previous burst as measured peak-to-peak, giving a burst rate of $\brate = 1 / \dt$.

Our inclusion of a nuclear heat source during initialisation of the envelope had substantially reduced thermal burn-in (\S~\ref{subsec:method_kepler}).
Despite this improvement, the first burst still ignites in a chemically pristine envelope, which lacks the complex ashes later accumulated.
Due to `chemical inertia', the models typically also require a `chemical burn-in' of several bursts to reach a quasi-stable bursting pattern.
To minimize these combined burn-in effects, we excluded the first 10 bursts of each model from analysis.
In previous studies, only $\approx 3$ had typically been discarded \citep{woosley_models_2004}.

Modelled and observed X-ray bursts are occasionally followed by short waiting-time bursts ($\dt \lesssim \SI{45}{\minute}$), which are thought to be triggered by the ignition of unburned hydrogen \citep{keek_thermonuclear_2017}.
These unusually weak bursts were not included in the data set of \citetalias{galloway_thermonuclear_2017}, and we excluded them from our analysis using the threshold of $\dt < \SI{45}{\minute}$.

We averaged the extracted properties of the remaining 20--30 bursts of each model, and adopted the $1 \sigma$ standard deviations as the model uncertainties.
Before averaging, each burst sequence was visually inspected to check there were no obvious secular trends in the properties (as illustrated in Figure~\ref{fig:preheating}).
This process produced tabulated burst properties across the model grid parameters (Fig.~\ref{fig:grid}), which could be sampled with minimal computational cost.

\subsection{Grid interpolation}
\label{subsec:method_interpolation}
Computational speed is critical for the large-scale sampling used in MCMC methods.
The smooth and monotonic behaviour of the burst properties (Fig.~\ref{fig:grid}) allowed for the use of interpolation between model predictions.
Using multivariate linear interpolation, we constructed a continuous function of the burst properties across the five grid parameters, $\mdot$, $\qb$, $\hyd$, $\cno$, and $g$.

Any point within the grid bounds of Table~\ref{tab:params} could then be quickly ($\ll \SI{1}{\second}$) sampled to predict the burst properties of interest: $\brate$, $\Lpeak$, and $\Eb$.
In contrast to the roughly $\SIrange{40}{90}{\hour}$ required for each \kepler{} simulation, this approach granted a considerable efficiency gain.

The interpolated burst properties were in the local neutron star frame of the \kepler{} models.
In order to compare with the observed data (Table~\ref{tab:obsdata}) we converted these values to observable quantities.
These calculations first accounted for the fact that \kepler{} uses Newtonian gravity.
A \kepler{} model with a given Newtonian surface gravity, $g$, can be considered equivalent to a neutron star with an equal $g$ under GR, but a different `true' mass and radius \citep[e.g.,][]{woosley_models_2004, keek_multi-zone_2011}.
In the interest of clarity, we will here use quantities that are already corrected for GR\footnote{These corrections are not applicable to codes which already use GR surface gravity, for example, \mesa}.
The basic corrections for Newtonian quantities are given in Appendix~\ref{sec:appendix_gr}.

\subsection{Free parameters: mass, distance, and anisotropy}
\label{subsec:method_free_params}
In addition to the grid parameters (Table~\ref{tab:params}), a set of `free' parameters are used for calculating observables.

For a given surface gravity, $g$, we can freely choose the neutron star mass, $M$, which determines the corresponding radius, $R$.
Both $M$ and $R$ then determine the gravitational redshift, $z$, which quantifies the neutron star compactness, given by
\begin{equation}
    \label{eq:redshift}
    z = \frac{1}{\sqrt{1 - 2 G M / (c^2 R)}} - 1,
\end{equation}
where $G$ is the gravitational constant and $c$ is the speed of light.
The distance to the system, $d$, is freely chosen.

Finally, we can choose the anisotropy factors, $\xib$ and $\xip$.
These parameters represent the deviation of the observed flux, $F$, from an isotropic neutron star luminosity, $L$, caused by the scattering and blocking of light by the accretion disc \citep[e.g.,][]{fujimoto_angular_1988, he_anisotropy_2016}.
The anisotropy parameters are defined with
\begin{equation}
    \label{eq:xi}
    \sub{F}{b} = \frac{\sub{L}{b}}{4 \pi d^2 \xib}, \quad \sub{F}{p} = \frac{\sub{L}{p}}{4 \pi d^2 \xip},
\end{equation}
where the subscripts `b' and `p' correspond to the burst and persistent emission, respectively.

Because $\xib$ and $\xip$ are degenerate with distance, we combined them into independent parameters: a modified distance, $\db$, and the anisotropy ratio, $\xiratio$.
We can later retrieve the absolute values for $\xib$, $\xip$, and $d$ by choosing an accretion disc model which relates the anisotropy to the system inclination, $i$ (\S~\ref{subsec:results_distance}).

\subsection{Transforming to observable quantities}
\label{subsec:method_observables}
The free parameters can then be used to calculate observables from the local burst properties.
We here signify observed quantities with the subscript `$\obssymb$'.

The burst rate as seen by a distant observer is time-dilated with
\begin{equation}
    \label{eq:rate_dt}
     \obs{\brate} = \frac{\brate}{1 + z}.
\end{equation}
The observed peak flux is given by
\begin{equation}
    \label{eq:peak}
    \obssub{F}{peak} = \frac{\Lpeak}{4 \pi d^2 \xib (1+z)^2}.
\end{equation}
The observed fluence is given by
\begin{equation}
    \label{eq:fluence}
    \obssub{f}{b} = \frac{\Eb}{4 \pi d^2 \xib (1+z)},
\end{equation}
where we use the redshift factor of $1+z$ instead of $(1+z)^2$, because fluence is time-integrated.

The two remaining observables, $\Fedd$ and $\Fper$, are not predicted from the model grid, but are calculated analytically from the given parameters.

The observed Eddington flux is given by
\begin{equation}
    \label{eq:fedd}
    \obssub{F}{Edd} = \frac{\Ledd}{4 \pi d^2 \xib (1+z)^2},
\end{equation}
where $\Ledd$ is the local Eddington luminosity, given by
\begin{equation}
    \label{eq:ledd}
    \Ledd = \frac{8 \pi G \sub{m}{p} c (1+z) M}{\sub{\sigma}{T} (\hyd + 1)},
\end{equation}
where $\sub{m}{p}$ is the mass of the proton and $\sub{\sigma}{T}$ is the Thompson scattering cross section.
This equation assumes that the radiation pressure is exerted on the electrons of an ionized plasma, and includes a scaling factor of $2 / (\hyd + 1)$ to account for the charge per mass for a given composition of hydrogen and helium.

The observed persistent flux is given by
\begin{equation}
    \label{eq:fper}
    \obssub{F}{p} = \frac{\Lper}{4 \pi d^2 \xip (1+z)^2},
\end{equation}
where $\Lper = -4 \pi R^2 \mdot \phi$ is the local accretion luminosity, and $\phi = - c^2 z/(1+z) \approx -0.2 \,c^2$ is the gravitational potential at the neutron star surface.

\subsection{Multi-epoch model}
\label{subsec:method_multi-epoch}
To model bursts from multiple epochs, we used our interpolated grid (\S~\ref{subsec:method_interpolation}) to predict the observed burst properties (\S~\ref{subsec:method_observables}) of three separate epochs from \gs{} (\S~\ref{subsec:method_observations}).
Our multi-epoch model contained both epoch-independent and epoch-dependent parameters.

The accreted composition and neutron star properties, $\hyd$, $\cno$, $g$, and $M$, are expected to remain unchanged between the observed epochs, and so global parameters are used.
Global parameters were also used for the distance and anisotropy, $\db$ and $\xiratio$, although the anisotropy factors $\xib$ and $\xip$ could feasibly evolve due to changes in the accretion disc.
We leave the testing of epoch-dependent parameters of $\xib$ and $\xip$ for a future study.
The six epoch-independent parameters are thus $\hyd$, $\cno$, $g$, $M$, $\db$, and $\xiratio$.

In contrast, the accretion rate is expected to evolve between epochs, and so we used three parameters, $\mdotn{1}$, $\mdotn{2}$, and $\mdotn{3}$, where the subscripts 1--3 correspond to the 1998, 2000, and 2007 epochs, respectively.
The crustal heating efficiency is predicted to depend on accretion rate \citep{cumming_long_2006}, and we similarly use three parameters for the flux emerging from the crust (i.e., ``base heating''), $\qbn{1}$, $\qbn{2}$, and $\qbn{3}$.
These were the first burst models to vary $\qb$ between accretion epochs.

The 12 parameters could then be separated into the grid parameters, $\mdotn{i}$, $\qbn{i}$, $\hyd$, $\cno$, and $g$, for $i = 1$, 2, 3, and the free parameters, $M$, $\db$, and $\xiratio$.
For a given point in parameter space, the former were used to interpolate the local burst properties from the grid, and the latter were used to transform these properties into observables.

We could then apply MCMC methods to draw samples from the multi-epoch model, and compare the predictions to observations to obtain probability distributions over the parameter space.

\subsection{MCMC methods}
\label{subsec:method_mcmc}
A detailed description of MCMC algorithms is beyond the scope of this paper, but many introductions are available \citep[e.g.,][]{mackay_information_2003}, and we provide here a brief summary.

The target distribution to be sampled -- the posterior probability distribution, or simply the `posterior' -- represents the probabilities over model parameters, given a set of data we wish to model.
The posterior is given by
\begin{equation}
    \label{eq:posterior}
    \posterior = \frac{1}{Z} \likelihood \prior,
\end{equation}
where $\theta$ represents the model parameters and $D$ is the data to be modelled.
The likelihood function, or simply the `likelihood', $\likelihood$, represents the probability of observing the data given the model predictions.
The prior distribution, or simply the `prior', $\prior$, represents any existing beliefs or constraints on the parameters.
The normalisation constant, $Z = p(D)$, also known as the `evidence', is independent of $\theta$, and the MCMC algorithm can draw samples from $\posterior$ without calculating $Z$.

An MCMC simulation consists of an ensemble of `walkers' in parameter space.
For each walker, the likelihood is evaluated at its location, and a new step is randomly drawn from a proposal distribution, typically a Gaussian centred on the walker.
The likelihood is then evaluated at the proposed step, which is either accepted or rejected based on the relative probability of the two points.
Through this repeated process, the walkers `explore' the parameter space of the posterior distribution.
After a sufficiently large number of steps, each point in the chain of steps represents an independent sample drawn from the posterior distribution.
The density of the walkers thus corresponds to the probability density.

A major advantage of MCMC methods is the ability to `marginalise' over uninteresting parameters,
by projecting the probability density onto a subset of dimensions.
We used marginalisation to produce 1D and two-dimensional (2D) distributions for the parameters of interest.

For our MCMC calculations, we used the open-source \python{} package \emcee\footnote{\url{https://emcee.readthedocs.io/en/v2.2.1/}} (V2.2.1), an affine-invariant ensemble sampler \citep{foreman-mackey_emcee:_2013}.
We constructed the posterior function defined in Eq.~\eqref{eq:posterior}, ignoring $Z$, and used \emcee{} to generate a chain of samples drawn from it.
This function takes the 12 multi-epoch model parameters (\S~\ref{subsec:method_multi-epoch}) as input, and calculates a posterior likelihood using the prior, $\prior$, and the model likelihood, $\likelihood$.

For the grid parameters, the limits were set by the boundaries of the model grid (Table~\ref{tab:params}).
For the free parameters, we imposed limits of $1.0 \leq M \leq \SI{2.2}{\msun}$, $1 \leq \db \leq \SI{15}{kpc}$, and $0.1 \leq \xiratio \leq 10$.
The prior distribution for each parameter was set to $\prior = 0$ outside these boundaries.

We used flat prior distributions for all parameters except $\cno$, setting $\prior = 1$ everywhere inside the boundaries.
For $\cno$, we estimated a prior distribution using a process similar to \citet{goodwin_bayesian_2019}.
From a simulated catalogue of Milky Way stars, constructed to represent the underlying distributions of the Gaia DR2 catalogue \citep{rybizki_gaia_2018}, we took a sample of $\num{100000}$ stars located within $\SI{15}{arcmin}$ of \gs, and between a distance of $\numrange{5}{9}\, \si{kpc}$.
We then fit a beta distribution to [Fe/H], obtaining the values of $\alpha = 10.1$ and $\beta = 3.5$ after translating to the interval $[-3.5,\, 1]$, which contained the vast majority of star samples.
We used this as the prior distribution for $\log_{10} (\cno / 0.01)$, where we have assumed a solar CNO metallicity of $0.01$ \citep{martienssen_4.4_2009}.
This distribution was applied inside the grid bounds of $0.0025 \leq \cno \leq 0.03$, which roughly corresponds to $-0.6 \leq \log_{10} (\cno / 0.01) \leq 0.5$.

For a given sample point in parameter space, $\theta$, the local burst properties were interpolated from the model grid (\S~\ref{subsec:method_interpolation}), from which the observables were predicted (\S~\ref{subsec:method_observables}).
The likelihood function, $\likelihood$, from Equation~\eqref{eq:posterior}, was then evaluated by comparing these predictions with the observed data, $D$, using
\begin{equation}
\label{eq:lhood}
    \likelihood = \prod_x \frac{1}{2 \pi (\sigma^2 + \sigma_\obslhood^2)} \exp{\left[ -\frac{(x - x_\obslhood)^2}{2 (\sigma^2 + \sigma_\obslhood^2)} \right]},
\end{equation}
where $x$ is iterated over each observable of each epoch, $\sigma$ is the uncertainty, and the subscript `\obslhood' signifies the corresponding observed value from Table~\ref{tab:obsdata}.

The MCMC model used an ensemble of $\num{1000}$ walkers, which were initialised in a small `hyperball' in parameter space.
The algorithm was run for $\num{20000}$ steps, resulting in a total of $\num{2e7}$ individual samples.
The average computation time for each sample was $\approx \SI{0.012}{s}$, for a total of $\approx 560$ CPU hours split over 8 cores.
For comparison, each \kepler{} simulation costs roughly 40--70 CPU hours.

We discarded the first \num{1000} steps as burn-in, after which the walkers had spread out across the domain of each parameter.
To check convergence, we estimated the autocorrelation time ($\tau$) at multiple steps in the chain\footnote{using \python{} code adapted from \url{https://dfm.io/posts/autocorr/}}, to ensure the total chain length was longer than $10\, \tau$.

%% file: sections/results.tex

\renewcommand{\arraystretch}{1.5}
\begin{table}
    \centering
    \caption{Maximum likelihood estimates for each 1D marginalised posterior, with 68~per~cent credible intervals.
	In addition to the twelve MCMC parameters (\S~\ref{subsec:method_multi-epoch}), we include the derived neutron star properties, $R$ and $z$ (\S~\ref{subsec:results_massradius}), the system properties predicted using a disc model of anisotropy, $i$, $\xib$, $\xip$, and $d$ (\S~\ref{subsec:results_distance}), and the global accretion rates, $\Mdotn{i}$ (\S~\ref{subsec:results_global_mdot}).
	The subscripts of 1--3 correspond to the 1998, 2000, and 2007 epochs, respectively.
	The accretion rates are given as fractions of the fiducial values, $\mdotedd = \SI{8.775e4}{g.cm^{-2}.s^{-1}}$ and $\Mdotedd = \SI{1.75e-8}{\msun.yr^{-1}}$.
	}
    \label{tab:max_lhood}
    \begin{tabular}{lll}
        \hline
		Parameter & Units & Estimate \\
		\hline
		$\mdotn{1}$ 			      & ($\mdotedd$)			& $0.083^{+0.013}_{-0.011}$ \\
		$\mdotn{2}$ 				  & ($\mdotedd$)			& $0.114^{+0.016}_{-0.017}$ \\
		$\mdotn{3}$ 				  & ($\mdotedd$)			& $0.132^{+0.018}_{-0.02}$ \\
		$Q_\mathrm{b,1}$ 			  & ($\si{\mevnuc}$)  		& $0.36^{+0.10}_{-0.2}$ \\
		$Q_\mathrm{b,2}$ 			  & ($\si{\mevnuc}$)  		& $0.17^{+0.10}_{-0.14}$ \\
		$Q_\mathrm{b,3}$ 			  & ($\si{\mevnuc}$)  		& $0.15^{+0.1}_{-0.11}$ \\
		$\hyd$ 						  & (Mass fraction) 		& $0.74^{+0.02}_{-0.03}$ \\
		$\cno$			 			  & (Mass fraction)   		& $0.010^{+0.005}_{-0.004}$ \\
        $g$				 		      & ($\SI{e14}{cm.s^{-2}}$) & $2.8^{+0.4}_{-0.6}$ \\
		$M$				    	      & ($\si{\msun}$)			& $>1.7$ \\
		$d \sqrt{\xi_\mathrm{b}}$  	  & ($\si{kpc}$)		   	& $6.5^{+0.4}_{-0.6}$ \\
		$\xi_\mathrm{p} / \xi_\mathrm{b}$ & -- 					& $1.57^{+0.15}_{-0.19}$ \\

		$R$ 						  & (km)                    & $11.3\pm 1.3$ \\
		$z$                           & --						& $0.39^{+0.07}_{-0.07}$ \\
		
		$i$ 						  & (deg) 					& $69^{+2}_{-3}$ \\
		$\xi_\mathrm{b}$ 			  & -- 						& $1.22^{+0.05}_{-0.06}$ \\
		$\xi_\mathrm{p}$ 			  & -- 						& $2.0^{+0.2}_{-0.4}$ \\
		$d$ 						  & (kpc) 					& $5.8^{+0.3}_{-0.4}$ \\

		$\Mdotn{1}$ 				  & ($\Mdotedd$)            & $0.098^{+0.012}_{-0.014}$ \\
		$\Mdotn{2}$ 				  & ($\Mdotedd$)            & $0.132^{+0.016}_{-0.02}$ \\
		$\Mdotn{3}$ 				  & ($\Mdotedd$)            & $0.153^{+0.017}_{-0.02}$ \\
		\hline
    \end{tabular}
\end{table}

\begin{figure*}
	\centering
	\includegraphics[width=\textwidth]{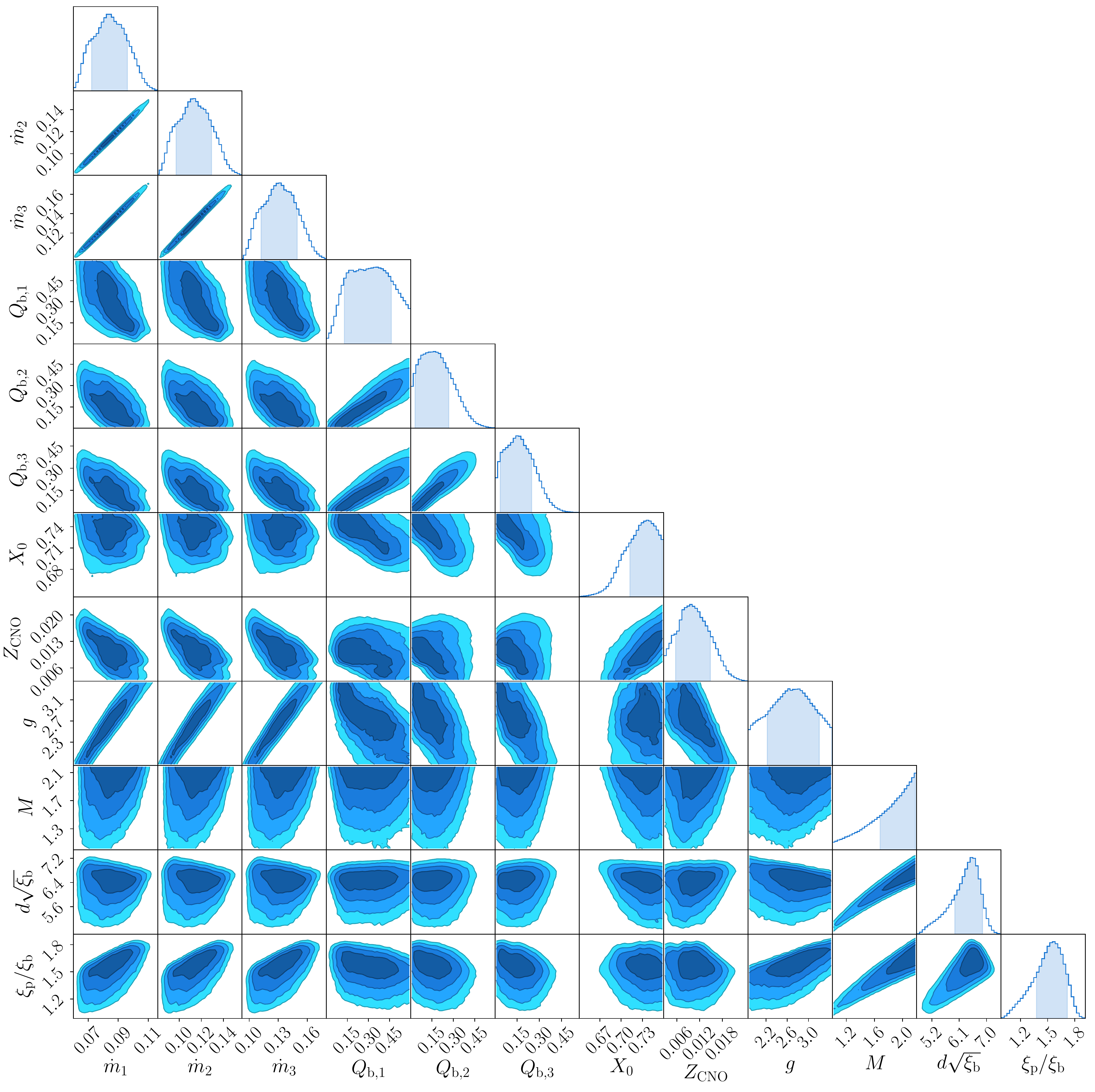}
	\caption{Posterior distributions for all twelve parameters of the multi-epoch MCMC model.
	The 2D contour levels indicate the 38, 68, 87, and 95~per~cent credible regions.
	Along the diagonal are the 1D marginalised posteriors, with the 68~per~cent credible interval shaded.
	The maximum likelihood estimates for the 1D posteriors are summarised in Table~\ref{tab:max_lhood}, along with the units for each parameter.
	}
	\label{fig:corner}
\end{figure*}

We present here the distributions and estimates from the burst matching procedure, and discuss the implications for the system properties.
The 2D marginalised posteriors for $\mdotn{i}$, $\qbn{i}$, $\hyd$, and $\cno$, are shown in Fig.~\ref{fig:corner}, with the 1D posteriors shown along the diagonal.
The maximum likelihood estimates for the 1D posteriors are listed in Table~\ref{tab:max_lhood}.
Unless otherwise stated, the uncertainties given for 1D parameter estimates are 68~per~cent credible intervals, and the 2D contour levels are 38, 68, 87, and 95~per~cent credible regions.

There is a strong correlation visible between the accretion rates of each epoch, and the base heating of each epoch.
These correlations are expected, because for given ratios between the epoch burst properties,  similar ratios are required between the epoch parameters.
For example, the persistent flux, $\Fper$, is calculated using Eq.~\eqref{eq:fper}, and is proportional to $\mdot$.

The CNO mass fraction, $\cno$, is correlated with the hydrogen fraction $\hyd$ -- a common feature of such model-observation comparisons \citep[e.g.,][]{galloway_helium-rich_2006, goodwin_bayesian_2019}.
The correlation arises because multiple pairs of $X_0$ and $\cno$ result in the same reduced hydrogen fraction at ignition.

The distributions of some parameters, for example $\qbn{1}$, $\hyd$, $M$, and $\cno$, appear to be truncated by the prior boundaries.
These limits could bias the results, potentially underestimating the full extent of the distributions.
Some of these boundaries were chosen as expected natural limits, whereas others are simply limited by the model grid.
For example, $\hyd$ is truncated at the upper grid limit of $\hyd=0.76$.
Although models with larger $\hyd$ could be added, they would substantially exceed the primordial mass fraction from Big Bang nucleosynthesis \citep{makki_critical_2019}.

On the other hand, $Q_\mathrm{b,1}$ appears truncated at the upper grid limit of $\SI{0.6}{\mevnuc}$.
This value is larger than the typical assumed heating of $\approx \SI{0.15}{\mevnuc}$ \citep[e.g.,][]{heger_models_2007}, although the amount of crustal heating emerging into the envelope is poorly constrained, and a total of 1--2 $\si{\mevnuc}$ is potentially available \citep{haensel_models_2008}.
The CNO metallicity, $\cno$, is also slightly limited by the lower grid boundary of $\cno = 0.0025$.
A future study could extend the model grid in these parameters, and examine the effect on the posteriors.

\begin{figure}
  \centering
  	\includegraphics[width=\linewidth]{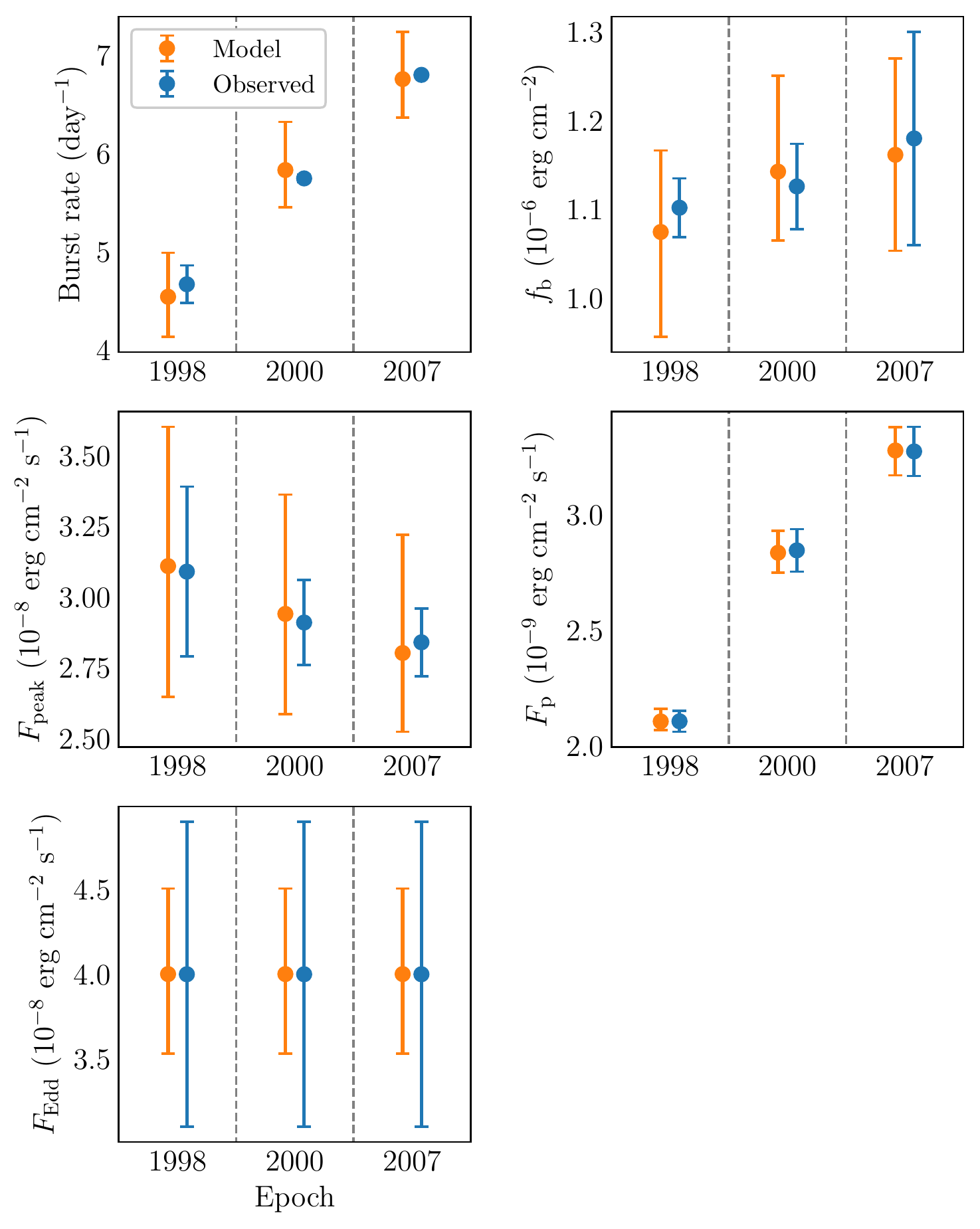}
	\caption{The distributions of the model-predicted observables from the MCMC chain (i.e.,\ the posterior predictive distribution;
	orange points), plotted against the observed data (blue points).
	The central points are the distribution peaks, and the error bars are 68~per~cent intervals.
	Each sample of the MCMC chain predicts these observables using the multi-epoch model (\S~\ref{subsec:method_multi-epoch}).
	The observed data are consistent with all predicted distributions, within uncertainties.
	}
	\label{fig:compare}
\end{figure}

\subsection{Predicted observables}
\label{subsec:results_observables}
The distribution of burst properties predicted over the MCMC simulation corresponds to the posterior predictive distribution.
This distribution represents the expected observations according to our model, given the posteriors of the model parameters.
A random sample of \num{20000} points were selected from the chain, and the predicted triplets of observables were extracted for each point using the multi-epoch model (\S~\ref{subsec:method_multi-epoch}).
The distribution peaks and 68~per~cent credible intervals are plotted with the original observed data in Fig.~\ref{fig:compare}.
The predicted burst properties are consistent with the observed data, within uncertainties.
Note that this comparison should not be confused with the `best-fit', which MCMC methods are ill-suited to finding.

\subsection{Crustal heating and accretion rate}
\label{subsec:results_qb_mdot}
By using independent base heating rates between epochs, we can examine the constraints on $\qb$ as a function of $\mdot$.
Theoretical models predict that the effective crustal heating is stronger at low accretion rates, and weaker at higher accretion rates due to neutrino losses \citep{cumming_long_2006}.

\begin{figure}
  \centering
  	\includegraphics[width=\linewidth]{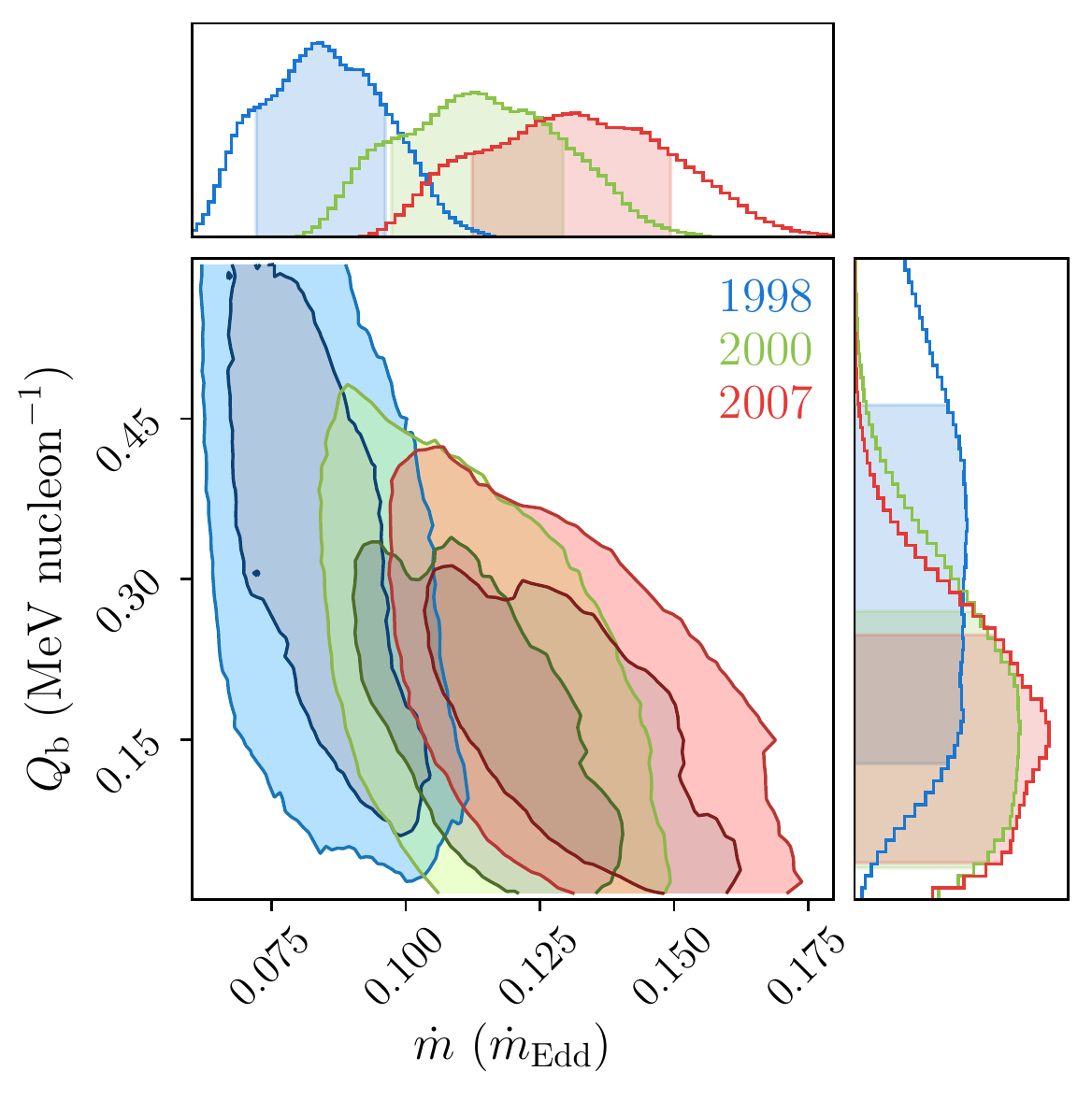}
	\caption{2D posteriors for the base heating and accretion rate, separated by epoch.
	For clarity, only the 68 and 95~per~cent contours are shown.
	There is a slight suggestion of an anti-correlation between base heating (and thus crustal heating) and accretion rate, although the uncertainty regions are large and overlapping.
	}
	\label{fig:qb_mdot}
\end{figure}

The posteriors of $\qb$ and $\mdot$ for each epoch are plotted in Fig.~\ref{fig:qb_mdot}.
There is significant overlap between the distributions, particularly between the 2000 and 2007 epochs, which have similar estimates for $\qb$.
The 1998 epoch, with the lowest inferred $\mdot$, covers similar $\qb$ but is overall consistent with larger values, with a 68~per~cent 2D credible region extending up to the grid boundary of $\SI{0.6}{\mevnuc}$, compared to $\approx \SI{0.3}{\mevnuc}$ for 2000 and 2007.

This comparison, though inconclusive, suggests an anticorrelation between $\qb$ and $\mdot$, as expected, but further investigation is needed.
Modelling burst epochs which span a larger range of accretion rates could help to constrain this relationship.

\begin{figure}
  \centering
  	\subfloat{\includegraphics[width=0.5\linewidth]{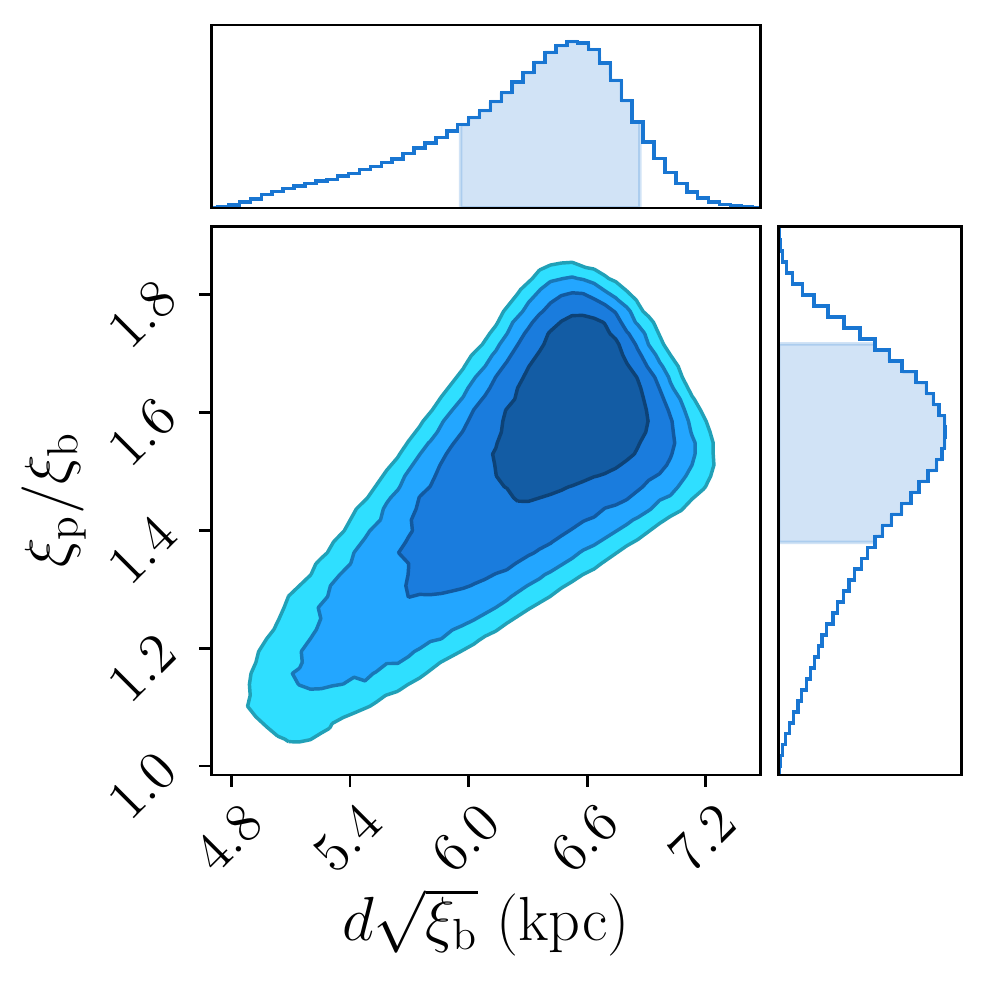}}
  	\subfloat{\includegraphics[width=0.5\linewidth]{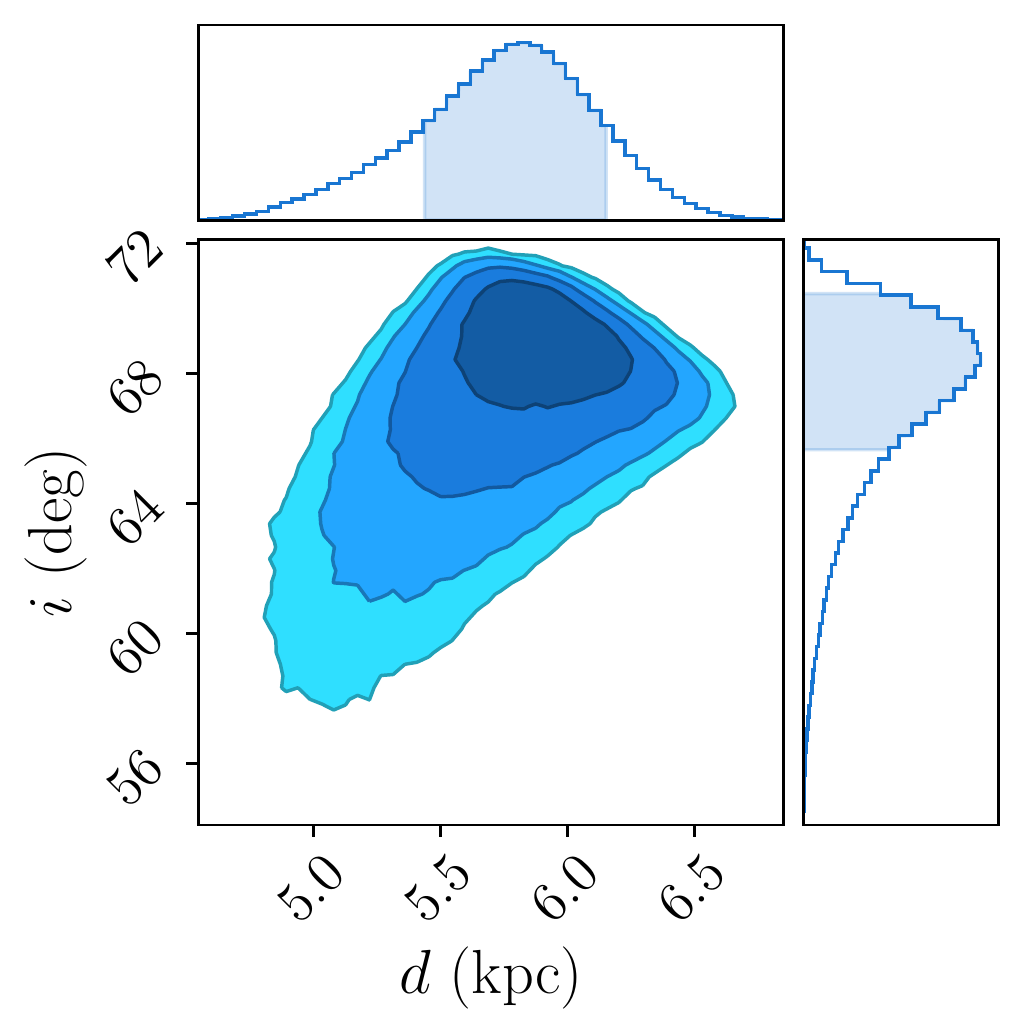}}
	\caption{The distance and anisotropy parameters from the MCMC simulation (left panel), and the  system inclination and absolute distance (right panel), predicted using the anisotropy model \textit{Disc a} from \citet{he_anisotropy_2016}.
	The contour levels and shaded regions are the same as Fig.~\ref{fig:corner}}
	\label{fig:inclination}
\end{figure}

\subsection{Distance and inclination}
\label{subsec:results_distance}
From $\xiratio$ we can derive constraints on the system inclination by adopting a disc model for anisotropy.
Disc models have been presented by \citet{he_anisotropy_2016}, which predicted the anisotropy according to the system inclination for multiple disc morphologies.
We used their model of a thin, flat disc (\textit{Disc a}) to predict the inclination, $i$, using $\xiratio$.
The disc model also predicts $\xip$ and $\xib$, from which we could obtain the absolute distance, $d$.
The posteriors for these quantities are plotted in Fig.~\ref{fig:inclination}, and the maximum likelihood estimates are listed in Table~\ref{tab:max_lhood}.

These estimates depend on the assumptions of the thin disc model, and only flat priors were used for $\db$ and $\xiratio$.
Exploring other priors, and other disc models, could yield different constraints.

\begin{figure}
  \centering
  	\includegraphics[width=\linewidth]{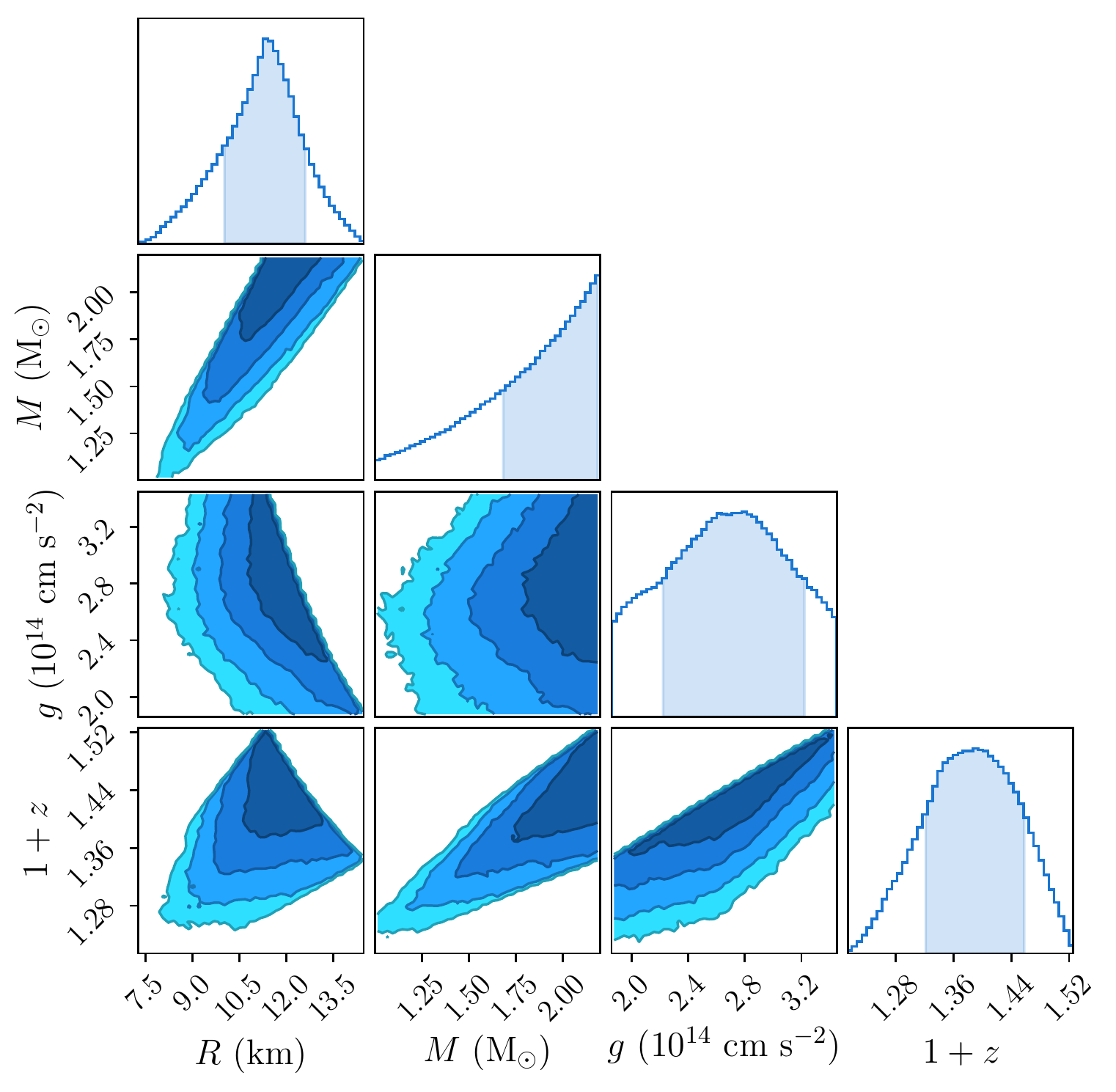}
	\caption{Posteriors for the gravitational parameters.
	The contour levels and shaded credible intervals are the same as Fig.~\ref{fig:corner}.
	The mass and surface gravity are parameters of the MCMC model, whereas the radius and redshift are derived from these quantities (\S~\ref{subsec:results_massradius}).
	The sharp boundaries visible in multiple contours correspond to the upper limit of $M$.
	}
	\label{fig:gravitational}
\end{figure}

\subsection{Neutron star properties}
\label{subsec:results_massradius}
We extract distributions for the neutron star properties using the MCMC parameters of $M$ and $g$.
The neutron star radius is calculated by solving
\begin{equation}
    g = \frac{G M}{R^2 \sqrt{1 - 2 G M / (c^2 R)} }
\end{equation}
for $R$, given $M$ and $g$.

The gravitational redshift, $z$, is then calculated from $M$ and $R$ using Eq.~\eqref{eq:redshift}.
The posteriors for these quantities are plotted in Fig.~\ref{fig:gravitational}, and the maximum likelihood estimates are listed in Table~\ref{tab:max_lhood}.

The highest probability density for $M$ is against the upper boundary of $M=\SI{2.2}{\msun}$, indicating that the distribution is truncated.
This upper limit was informed by the maximum neutron star mass of $\approx \SI{2.17}{\msun}$ inferred from the neutron star merger GW170817 \citep{margalit_constraining_2017}, suggesting a possible bias in our model towards large masses.
Additionally, the distribution for $g$ is constrained by both the upper and lower model grid boundaries.

We note that only flat prior distributions were used for $M$ and $g$, and thus did not include any expectations from theoretical EOS predictions, or from mass estimates of similar bursting systems \citep[e.g.,][]{ozel_mass_2012}.
Exploring additional prior distributions, and expanding the model grid in $g$, is required before drawing stronger conclusions.

Despite the limitations, these results represent a step towards constraining the neutron star mass and radius using 1D burst models.

\subsection{Global accretion rate}
\label{subsec:results_global_mdot}
The model grid uses the local accretion rate per unit area, $\mdot$.
The global accretion rate, $\Mdot = 4 \pi R^2 \mdot$, depends on the neutron star radius, which is determined by $g$ and the free parameter of $M$.
Combining the posterior samples of $\mdotn{i}$ and $R$, we obtain estimates for $\Mdotn{i}$, which are listed in Table~\ref{tab:max_lhood}.
We give $\Mdotn{i}$ as a fraction of the fixed Eddington rate, $\Mdotedd = \SI{1.75e-8}{\msun.yr^{-1}}$, which is the equivalent of $\mdotedd = \SI{8.775e4}{g.cm^{-2}.s^{-1}}$ for $R = \SI{10}{km}$.
We note again that this Eddington value is simply used as common reference points for convenience, and does not represent the `true' Eddington limit.

\subsection{Lightcurve sample}
\label{subsec:results_sample}
The burst data used by the MCMC model is an incomplete description of the full burst lightcurve.
The two quantities extracted from the lightcurves were the fluence, $\fluence$, and the peak flux, $\Fpeak$.
To test whether the full lightcurves of \kepler{} models remain consistent with the observations, we performed an additional set of simulations.

We took a random sample of 30 points from the MCMC chain, and for each point generated three new \kepler{} models using the sampled parameters $\mdotn{i}$, $\qbn{i}$, $\hyd$, $\cno$, and $g$.
The result was a set of 90 \kepler{} simulations, representing a sample of 30 epoch triplets from the posterior distribution.

\begin{figure}
  \centering
  	\includegraphics[width=\linewidth]{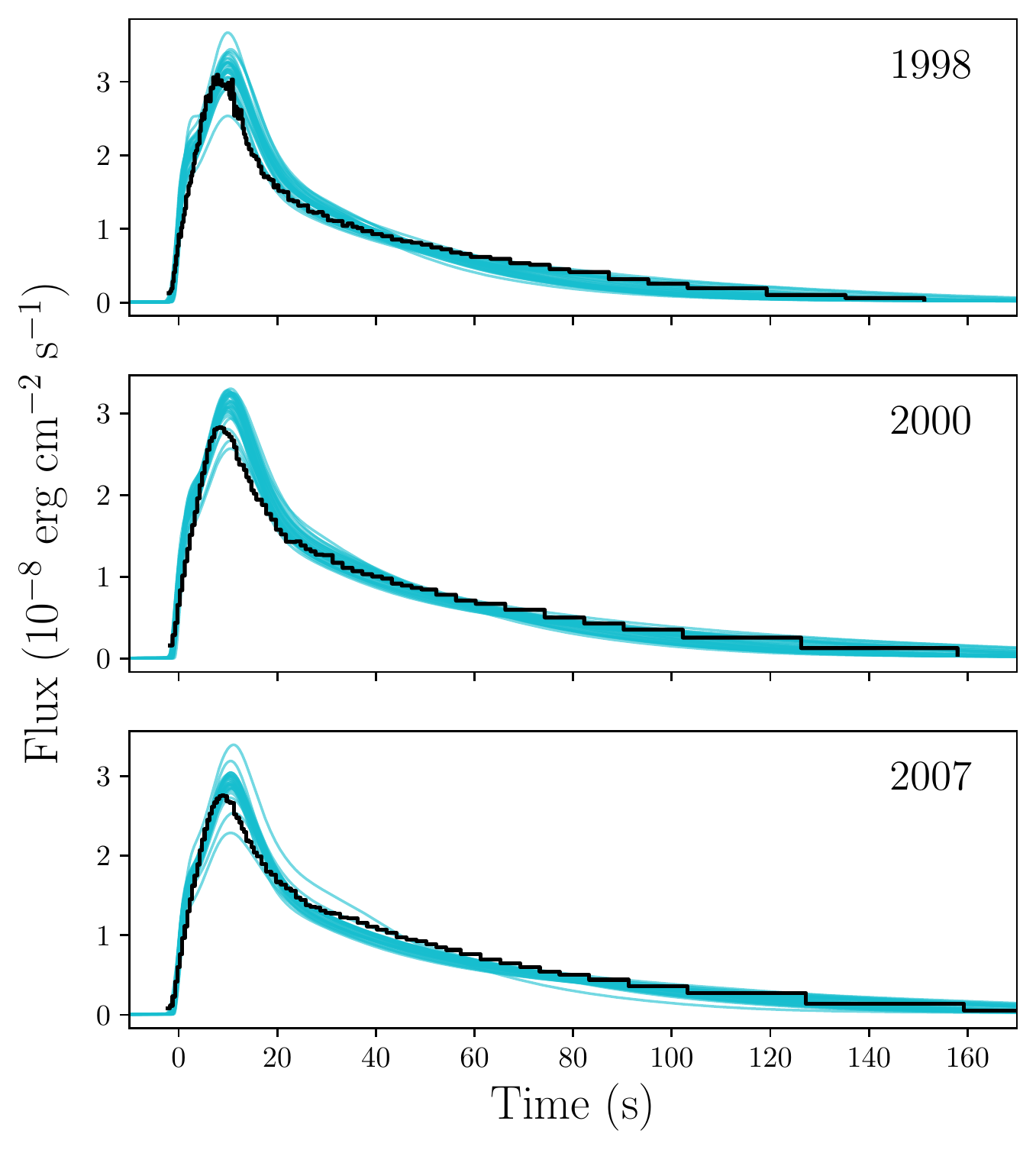}
	\caption{A comparison of the average lightcurves from 30 triplets of \kepler{} simulations (blue curves, 90 models total) and the observed epoch lightcurves (black histograms).
	The parameters for these additional \kepler{} models were drawn from 30 random samples of the MCMC chain.
	There is overall good agreement between the modelled and observed lightcurves, despite the MCMC procedure fitting only the fluence and peak flux from the observed lightcurve.
	For clarity, only the average curves are plotted, and the $1\sigma$ variations cover a wider range, as reflected by the $\Fpeak$ values in Fig.~\ref{fig:compare}.
	}
	\label{fig:lightcurve}
\end{figure}

The modelled bursts were extracted using the same procedure as the original grid (\S~\ref{subsec:method_extracting}).
We calculated average burst lightcurves for each model, and converted them to observable fluxes using the corresponding samples of $M$ and $\db$ (\S~\ref{subsec:method_observables}).
These lightcurves are plotted with the observed lightcurves in Fig.~\ref{fig:lightcurve}.

There is good agreement between the modelled and observed lightcurves, particularly considering that the MCMC model was fitting the fluence and peak flux, and not the full lightcurves.
This comparison suggests that these scalar quantities may be sufficient proxies for the overall lightcurve -- at least for bursts with similar lightcurve morphologies.

Nevertheless, some lightcurve information is still lost with this method.
For example, the morphology of the tail encodes further information about the \textit{rp}-process, the cooling of the envelope \cite{int_zand_long_2009}, and possible interactions between the burst flux and the disc \citep{worpel_evidence_2015}.
The `convexity` of the burst rise can also be calculated \citep{maurer_ignition_2008}, although 1D models cannot predict the effects of rotation, ignition latitude, and flame-spreading on this quantity \citep[e.g.,][]{zhang_link_2016}.
Fitting additional lightcurve data, or even the entire lightcurve itself (\S~\ref{sec:discussion}), should remain a goal for future model comparisons.

\begin{figure}
  \centering
  	\includegraphics[width=\linewidth]{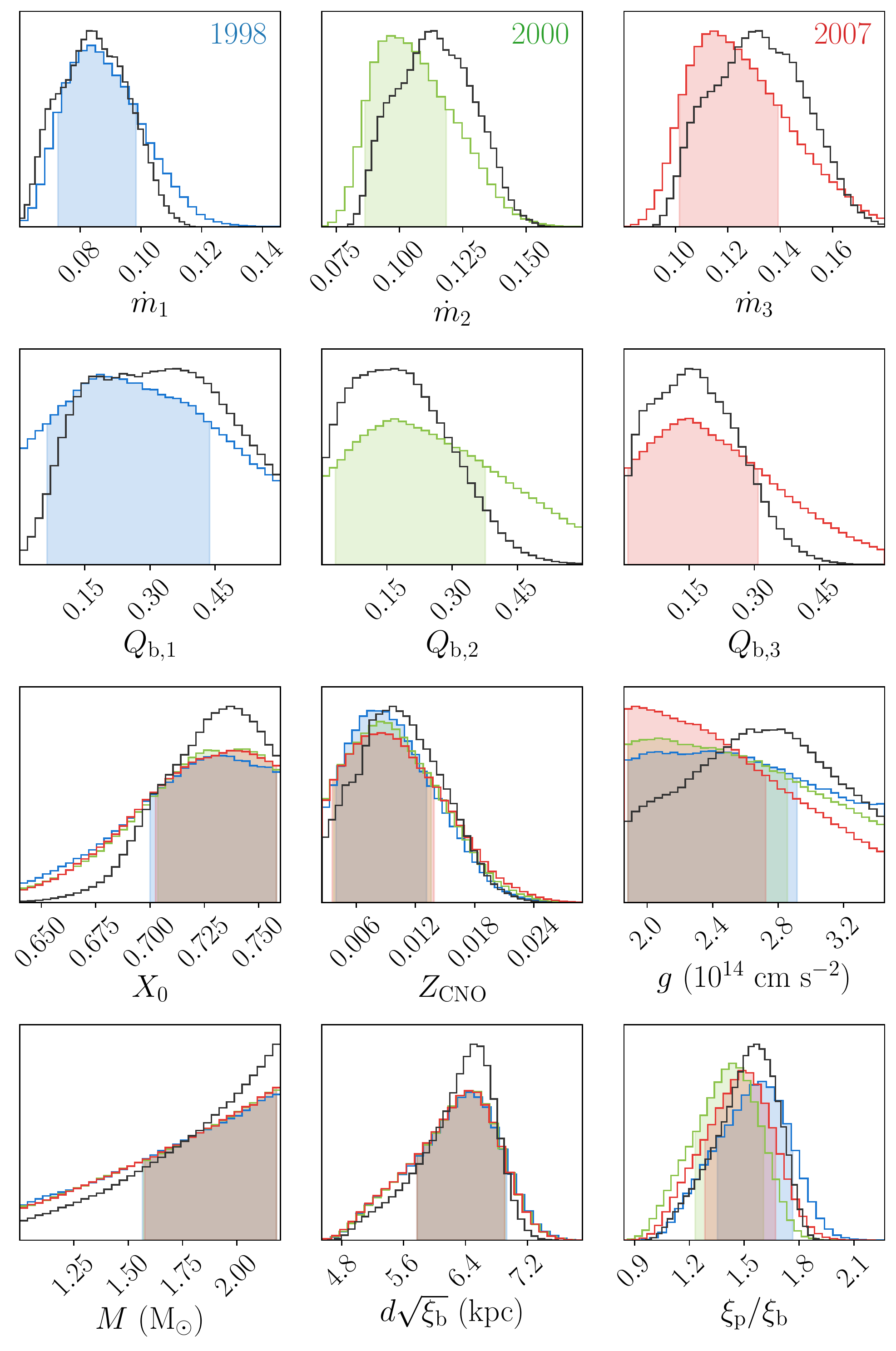}
	\caption{The posteriors for three additional MCMC models, each fitting only one epoch (coloured histograms), along with the original multi-epoch model (black histogram).
	The shaded regions are 68~per~cent credible intervals.}
	\label{fig:epoch_posteriors}
\end{figure}

\subsection{Modelling single epochs}
\label{subsec:results_epochs}
To test the benefit of fitting multiple epochs simultaneously, we performed three additional MCMC models, each fitting the data of a single epoch.
The posteriors for the single-epoch chains are shown in Fig.~\ref{fig:epoch_posteriors} (coloured histograms), along with the original multi-epoch posteriors (black histograms).

Compared to fitting the epochs separately, the posterior distributions were generally more constrained when all three epochs were fit simultaneously.
An exception appears to be $g$, although all four distributions for this parameter are heavily truncated at the boundaries, potentially interfering with the results.
The parameter constraints also remain overall consistent between the multi-epoch and single-epoch chains.

This comparison supports the approach that fitting multiple epochs simultaneously can help to improve the degeneracies between the system parameters \citepalias[as tested by][]{meisel_consistent_2018}.

%% file: sections/discussion.tex
We constrained system parameters for \gs{} by comparing multi-epoch observations to the most extensive set of 1D model predictions to date.
All central values discussed here correspond to the maximum likelihood estimates of the 1D marginalised posteriors, with 68~per~cent credible intervals (Table~\ref{tab:max_lhood}).

The global accretion rate estimates of $\Mdot = 0.098^{+0.012}_{-0.014}$, $0.132^{+0.016}_{-0.02}$, and $0.153^{+0.017}_{-0.02}\, \Mdotedd$, are roughly double those initially suggested by \citetalias{galloway_thermonuclear_2017} of $0.0513$, $0.0692$, and $0.0796\, \Mdotedd$, respectively, although those estimates did not account for anisotropy or different values of gravity.
Conversely, our central values are only $\approx 10$~per~cent smaller than those reported by \citetalias{meisel_consistent_2018} of $\mdot = 0.11,\, 0.15,$ and $0.17\, \mdotedd$, and are consistent within $2\sigma$.
Planned comparisons of \mesa{} and \kepler{} models could test whether their predictions are now more consistent, given the improvements to \kepler{} described in \S~\ref{subsec:method_kepler}.

The base heating estimates for the 2000 and 2007 epochs of $0.17^{+0.10}_{-0.14}$ and $0.15^{+0.1}_{-0.11}\, \si{\mevnuc}$ are centred near the canonical value of $\qb = \SI{0.15}{\mevnuc}$, though with broad credible intervals.
On the other hand, the estimate for the 1998 epoch, with a lower accretion rate (\S~\ref{subsec:results_qb_mdot}), is roughly double, at $0.36^{+0.10}_{-0.2}\, \si{\mevnuc}$, although $\SI{0.15}{\mevnuc}$ still lies within $1 \sigma$.
In agreement with \citetalias{meisel_consistent_2018}, base heating above $\qb \approx \SI{0.5}{\mevnuc}$ is disfavoured in all epochs, although the upper limit of our model grid is $\qb = \SI{0.6}{\mevnuc}$, compared to the $\SI{1.0}{\mevnuc}$ considered by \citetalias{meisel_consistent_2018}.

The CNO metallicity of $\cno = 0.01^{+0.005}_{-0.004}$ is centred on the assumed solar value of 0.01.
The chosen prior distribution was also centred near 0.01
(\S~\ref{subsec:method_mcmc}).
The result broadly supports a solar metallicity, whereas previous studies typically used higher values of $\cno = 0.02$ \citepalias[e.g.,][]{meisel_consistent_2018, heger_models_2007}.
Values below $\approx 0.005$ are disfavoured, such as the low-metallicity of $\cno = 0.001$ suggested by \citetalias{galloway_periodic_2004}.
Our model grid, however, only extends down to $\cno = 0.0025$, and could be expanded in future studies.

The accreted hydrogen fraction of $\hyd = 0.74^{+0.02}_{-0.03}$ is larger than the commonly-assumed value of $\hyd = 0.7$, which lies slightly outside $1 \sigma$, but still within $2 \sigma$ of our estimate.
The $1 \sigma$ credible interval extends up to $\hyd \approx 0.76$, at odds with \citetalias{meisel_consistent_2018}, who reported poor model fits for $\hyd=0.75$.

Studies which do not account for burst anisotropy in their distance estimates are implicitly reporting $\db$.
Our value of $\db = 6.5^{+0.4}_{-0.6}\, \si{kpc}$ is consistent with the previous estimates of $\SI{6}{kpc}$ \citepalias{meisel_consistent_2018}, $\SI{6.1}{kpc}$ \citepalias{galloway_thermonuclear_2017}, and $\SI{6.07 \pm 0.18}{kpc}$ \citepalias{heger_models_2007}.
Our distance is larger than the estimate of $\SI{5.7 \pm 0.2}{kpc}$ from \citet{chenevez_soft_2016}, which was obtained using the same 2010 $\Fedd$, but assumed a fixed mass of $M = \SI{1.4}{\msun}$ and a radius of $R = \SI{10}{km}$.
Our distance is also consistent with earlier upper limits of $\SI{8}{kpc}$ \citep{in_t_zand_broad-band_1999} and $7.5 \pm 0.5\, \si{kpc}$ \citep{kong_x-ray/optical_2000}, but is larger than the more recent upper limit of $\numrange{4.0}{5.5}\, \si{kpc}$ \citep{zamfir_constraints_2012}.

The anisotropy ratio of $\xiratio = 1.57^{+0.15}_{-0.19}$ agrees with the original estimate of $\xiratio = 1.55$ from \citetalias{heger_models_2007}, but not with the value of $3.5$ from \citetalias{meisel_consistent_2018}, although both of these studies explored fewer model parameters.
Using a flat disc model from \citet{he_anisotropy_2016}, we obtained from $\xiratio$ a system inclination of $i = {69^{+2}_{-3}}^{\circ}$.
This inclination is consistent with the upper limit of $i \lesssim 70^{\circ}$ from low-amplitude optical modulations \citep{homer_evidence_1998}, and the range of 40--70$^{\circ}$ suggested by \citet{mescheryakov_optical_2004} from models of the disc size.

The gravitational redshift of $z = 0.39 \pm 0.07$ is larger than the commonly-assumed value of $z=0.26$ from $M = \SI{1.4}{\msun}$ and $R = \SI{11.2}{km}$, and agrees with the value of $0.42$ from \citetalias{meisel_consistent_2018}, but is outside the inferred range of $z = \numrange{0.19}{0.28}$ for \gs{} reported by \citet{zamfir_constraints_2012}.
The radius of $R = 11.3\pm \SI{1.3}{km}$ is consistent with the rough upper limit of $\numrange{9.0}{13.2}\, \si{km}$ suggested by \citet{zamfir_constraints_2012}.

This study is the first application of MCMC methods to 1D burst models featuring adaptive nuclear networks.
As such, simplifying assumptions have been made and some care should be taken when interpreting the results.

It should be emphasized that our posterior statistics are fully dependent on the assumptions contained in the \kepler{} models and our interpolation between their predictions.
Although it is currently among the most advanced codes for simulating X-ray bursts, \kepler{} is still the subject of ongoing refinements (\S~\ref{subsec:method_kepler}), and is inherently limited to spherical symmetry.
Additionally, using linear interpolation to sample between the models  (\S~\ref{subsec:method_interpolation}) may introduce artificial `kinks' at the grid points, potentially affecting the resulting distributions.
Our comparisons of the posterior predictive distribution to the observed data (\S~\ref{subsec:results_observables}), and a sample of full lightcurves (\S~\ref{subsec:results_sample}), however, suggest that the interpolated models are not behaving unexpectedly.

The posterior distributions were truncated by some of the model grid boundaries, notably $\hyd$, $\cno$, $\qb$, and $g$ (Fig.~\ref{fig:corner}).
Some of these boundaries are physically-motivated, for example $\hyd < 0.76$ and $\qb > \SI{0}{\mevnuc}$, whereas others could realistically be extended, for example below $\cno = 0.0025$ and above $\qb = \SI{0.6}{\mevnuc}$.
Large extensions of the five-dimensional model grid, however, are limited by computational costs.

The information we included in our prior distributions (\S~\ref{subsec:method_mcmc}) was relatively limited.
All parameters except for $\cno$ used flat priors, which potentially give undue weight to physically unrealistic regions of parameter space.
For example, all combinations of $M$ and $g$ -- and by extension, the corresponding $R$ and $z$ -- were considered equally likely under the prior assumptions.
This may have contributed to the possible bias towards large $M$ (\S~\ref{subsec:results_massradius}).

The limitations discussed above can be investigated and improved upon in future work.
Linear interpolation, while computationally fast, has limited accuracy.
Other interpolation methods, such as cubic splines, could be explored, but care should be taken to avoid introducing artefacts.
A parameter sensitivity study could also identify which grid parameters can afford fewer model points, reducing the total number of simulations required.

The observed values of $\Fpeak$ and $\fluence$ were taken from observed lightcurves.
The full burst lightcurves, however, encode additional information about the rates of heating and cooling, and the extent of \textit{rp}-process burning in the tail.
Fitting whole lightcurves could improve, or even significantly reshape, the posteriors.
Implementing this approach, however, poses certain challenges.
Whereas interpolating scalar quantities is straightforward, it is unclear how best to do so for lightcurves.
If the lightcurves significantly change in morphology, interpolation could introduce nonphysical features.
A possible alternative is to use machine learning to efficiently predict lightcurves between models, such as the methods recently applied to gravitational waveforms of neutron star mergers \citep{easter_computing_2019}.
Extra parametrizations of the lightcurve could also be used, by fitting curves to the burst tail \citep{int_zand_long_2009}.
Nevertheless, our test of a limited sample of full lightcurves (\S~\ref{subsec:results_sample}) suggests that $\fluence$ and $\Fpeak$ may still serve as reasonable representations of the lightcurve.

A key benefit of MCMC methods is their ability to efficiently handle large numbers of parameters.
Additional parameters not used in this work could also be explored.
For example, using epoch-dependent anisotropy ratios, $\xiratio$, could test for possible changes in the accretion disc properties between epochs.
When calculating the Eddington flux, $\Fedd$, we assumed that the hydrogen fraction was equal to the accreted fraction, $\hyd$, but expansion of the outer layers during PRE may expose deeper hydrogen-poor layers, increasing $\Fedd$.
This hypothesis could be tested by including the hydrogen composition for $\Fedd$ as an additional parameter.

%% file: sections/conclusion.tex
We carried out Markov chain Monte Carlo (MCMC) simulations to model multi-epoch X-ray bursts from \gs{}.
By precomputing a grid of 3840 \kepler{} models, we interpolated the predicted burst properties and efficiently sampled the model parameter space.
Applying the Bayesian framework of MCMC allowed us to systematically examine the relationships between the model parameters and the predicted burst properties.
We obtained probability distributions for the properties of \gs{}, including the accretion rates, base heating rates, accreted composition, and surface gravity.

This work represents the most comprehensive use of 1D models on a burst source to date.
We have explored model parameters which are often held constant in burst models, including the base heating, accreted hydrogen composition, surface gravity, the neutron star mass and radius, and the gravitational redshift.
By using epoch-dependent parameters of $\qb$, we have also tested the dependence of crustal heating on accretion rate (\S~\ref{subsec:results_qb_mdot}), suggesting a preference for stronger crustal heating at lower accretion rates.

Although we have focused on \gs, the methods presented here are applicable to other X-ray burst observations.
Once the model grids are established, they can also be reused for similar systems.
By incorporating new epoch data and expanding the grid parameters, we can analyse additional sources suggested by \citetalias{galloway_thermonuclear_2017}, such as the helium-burster, \fouru.
Preliminary work is already underway to model PRE bursts from \fouru with a new model grid, which we plan to present in a future publication.

This work demonstrates the largely uncharted potential of using 1D burst models for the parameter estimation of neutron star systems.

%% file: misc/acknowledgements.tex
The authors thank the referee for their constructive feedback which has helped to improve the manuscript, Andrew Casey for useful discussions, and both Jordan He and Laurens Keek for providing data tables of their disc anisotropy models.
This work was supported in part by the National Science Foundation under Grant No.\ PHY-1430152 (JINA Center for the Evolution of the Elements).
This paper uses preliminary analysis results from the Multi-INstrument Burst ARchive (MINBAR), which is supported under the Australian Academy of Science's Scientific Visits to Europe program, and the Australian Research Council's Discovery Projects and Future Fellowship funding schemes.
This research was supported in part by the Monash eResearch Centre and eSolutions-Research Support Services through the use of the MonARCH HPC Cluster.
This work was supported in part by Michigan State University through computational resources provided by the Institute for Cyber-Enabled Research.
This work was performed in part on the OzSTAR national facility at Swinburne University of Technology.
OzSTAR is funded by Swinburne University of Technology and the National Collaborative Research Infrastructure Strategy (NCRIS).
ZJ was supported by an Australian Government Research Training Program (RTP) Scholarship.
AH was supported by an ARC Future Fellowship (FT120100363).

\textit{Software:}  \textsc{astropy} \citep{robitaille_astropy_2013},
\textsc{chainconsumer} \citep{hinton_chainconsumer_2016}, 
\textsc{emcee} \citep{foreman-mackey_emcee:_2013},
\textsc{matplotlib} \citep{hunter_matplotlib_2007}, 
\textsc{numpy} \citep{van_der_walt_numpy_2011}, 
\textsc{pandas} \citep{mckinney-proc-scipy-2010}, 
\textsc{scipy} \citep{virtanen_scipy_2020}.

%% file: appendix/preheating.tex
Excessive burn-in can occur during simulations if nuclear heating, $\qnuc$, is neglected during the model setup phase (\S~\ref{subsec:method_kepler}).
Addressing the issue is not straightforward, however, because nuclear heating occurs throughout the envelope at difference rates, depending on the local conditions.
By contrast, the flux from crustal heating, $\qb$, is simply implemented as a lower boundary condition.
Predicting $\qnuc$ in advance is difficult prior to running the full simulation with a nuclear network.

We added a heat source during the setup of the envelope before the full burst simulation begins.
For a chosen $\qnuc$, the total heat flux is given by $\sub{F}{nuc} = \qnuc \mdot$,
which we distributed throughout the envelope as a Gaussian function, centred at a column depth of $y \approx \SI{8e7}{\yunits}$ with a spread of $\sigma \approx \SI{8e6}{\yunits}$.
For comparison, $\qb$ is implemented at the lower model boundary of $y \sim \SI{e12}{\yunits}$.

\begin{figure}
  \centering
  	\includegraphics[width=\linewidth]{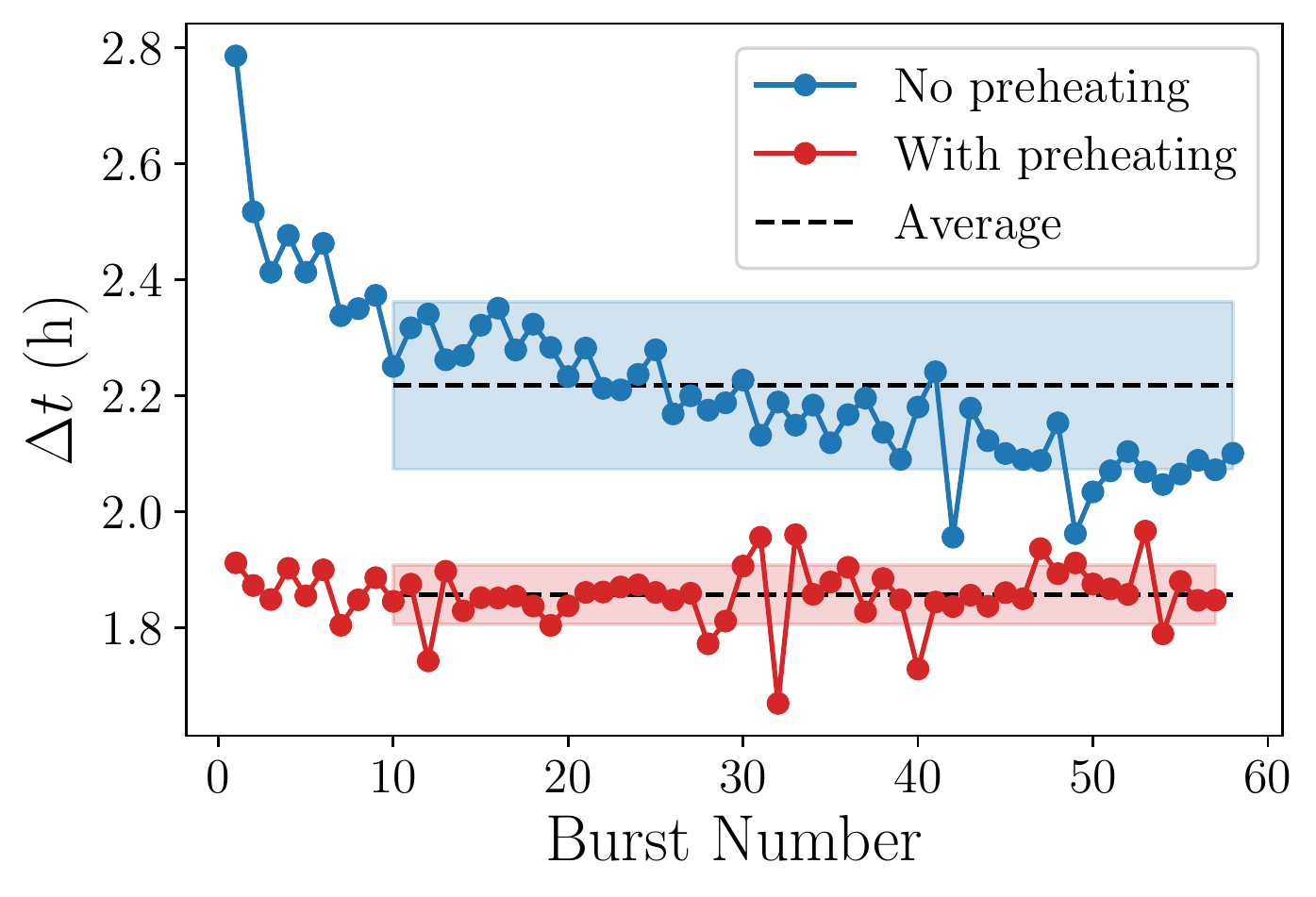}
	\caption{The burst train evolution for a simulation, with and without nuclear preheating.
	The dashed lines are the average values after discarding the initial ten bursts, and the shaded regions enclose the $1\sigma$ standard deviation.
	When nuclear preheating is included, the systematic burn-in is effectively eliminated, resulting in a smaller uncertainty and a consistent burst train.}
	\label{fig:preheating}
\end{figure}

We tested this model setup with a heating of $\qnuc = \SI{5}{\mevnuc}$, the approximate energy release expected for hydrogen burning.
We used model parameters of $\hyd = 0.73$, $\cno = 0.005$, $\mdot = 0.2$, and $\qb = \SI{0.05}{\mevnuc}$.
The burn-in was essentially eliminated from the resulting burst simulation (Fig.~\ref{fig:preheating}), in contrast to an identical model without preheating (effectively, $\qnuc = \SI{0}{\mevnuc}$).

Previous studies typically discarded only the first $\approx 3$ bursts to account for model burn-in \citep[e.g.,][]{woosley_models_2004}.
We demonstrate, however, that a 10--20~per~cent discrepancy can persist between $\dt$ even after 50 bursts.
Nevertheless, further investigation is required into the sensitivity of models to preheating.
Other bursting regimes, for example helium bursts, may require additional testing of the heating rates and depths.

%% file: appendix/gr.tex
\kepler{} uses Newtonian gravity to calculate the gravitational acceleration, given by
\begin{equation}
    g = \frac{G M}{R^2}.
\end{equation}
In the GR regime of a neutron star, however, the surface gravity is given by
\begin{equation}
    g = \frac{G M (1+z)}{R^2},
\end{equation}
where $z$ is the gravitational redshift given in Eq.~\eqref{eq:redshift}.

Because the envelope is a thin shell ($\Delta R \ll R$), the surface gravity is approximately constant throughout the envelope.
The Newtonian \kepler{} model is equivalent to neutron stars under GR with different $M$ and $R$, but with the same $g$.\
There is a contour of $M$ and $R$ pairs which satisfy this constraint.
For a chosen $M$ and $R$, the Newtonian \kepler{} quantities can be corrected to the equivalent GR values.
A more detailed description of these corrections can be found in Appendix~B of \citet{keek_multi-zone_2011}.

\citet{keek_multi-zone_2011} defined the mass and radius ratios between the two regimes,
\begin{equation}
    \label{eq:gr_xi_phi}
    \varphi = \frac{\gr{M}}{\nw{M}}, \quad \xi = \frac{\gr{R}}{\nw{R}},
\end{equation}
where we here signify the Newtonian and GR quantities with the subscripts `k', and `g', respectively.
Setting the requirement that $g$ must be equal under the two regimes, the above ratios are related by
\begin{equation}
    \label{eq:gr_redshift}
    \xi^2 = \varphi (1+z),
\end{equation}
where $z$ is evaluated for $\gr{M}$ and $\gr{R}$.

The ratio of the neutron star surface areas is given by $\xi^2$, and so the GR-corrected luminosity is given by
\begin{equation}
    \gr{L} = \xi^2 \nw{L} = \varphi (1+z) \nw{L}.
\end{equation}
For a given accretion rate per unit area, $\mdot$, the global accretion rate, $\Mdot = 4 \pi R^2 \mdot$, is also scaled by the area ratio,
\begin{equation}
    \gr{\Mdot} = \xi^2 \nw{\Mdot} = \varphi (1+z) \nw{\Mdot}.
\end{equation}
Both regimes are in the same local reference frame, and so $\dt$ and $\brate$ are not time-dilated.

These GR-corrected quantities were used to calculate the predicted observables in \S~\ref{subsec:method_free_params}~and~\ref{subsec:method_observables}.

%% file: appendix/data.tex
The data used in this work are publicly available in a Mendeley Data repository\footnote{\url{http://dx.doi.org/10.17632/nmb24z6jrp.1}}.
A table of the analysed model grid is included, listing the input parameters and output burst properties of each model as Newtonian \kepler{} quantities (i.e.,\ \textit{not} corrected for GR).
The full MCMC sample chains (including the discarded burn-in) are provided as 3D arrays of walkers $\times$ steps $\times$ parameters.
Further information on how to load this data is provided in the data repository.